\documentclass[
reprint,
superscriptaddress,
amsmath,amssymb,
aps,
]{revtex4-1}

\usepackage{color}
\usepackage{graphicx}
\usepackage[FIGTOPCAP]{subfigure}
\usepackage{dcolumn}
\usepackage{dcolumn}
\usepackage{bm,braket}

\begin {document}

\title{Sample-to-sample fluctuations of transport coefficients in the totally asymmetric simple exclusion process with quenched disorder}

\author{Issei Sakai}
\affiliation{%
    Department of Physics, Tokyo University of Science, Noda, Chiba 278-8510, Japan
}%


\author{Takuma Akimoto}
\email{takuma@rs.tus.ac.jp}
\affiliation{%
    Department of Physics, Tokyo University of Science, Noda, Chiba 278-8510, Japan
}%



\date{\today}

\begin{abstract}
    We consider the totally asymmetric simple exclusion processes on quenched random energy landscapes.
    We show that the current and the diffusion coefficient differ from those for homogeneous environments.
    Using the mean-field approximation, we analytically obtain the site density when the particle density is low or high.
    As a result, the current and the diffusion coefficient are described by the dilute limit
    of particles or holes, respectively.
    However, in the intermediate regime, due to the many-body effect, the current and the diffusion coefficient differ
    from those for single-particle dynamics.
    The current is almost constant and becomes the maximal value in the intermediate regime.
    Moreover, the diffusion coefficient decreases with the particle density in the intermediate regime.
    We obtain analytical expressions for the maximal current and the diffusion coefficient based on the renewal theory.
    The deepest energy depth plays a central role in determining the maximal current and the diffusion coefficient.
    As a result, the maximal current and the diffusion coefficient depend crucially on the disorder, i.e., non-self-averaging.
    Based on the extreme value theory, we find that sample-to-sample fluctuations of the maximal current and diffusion coefficient 
    are characterized by the Weibull distribution.
    We show that the disorder averages of the maximal current and the diffusion coefficient converge to zero as the system size is increased 
    and quantify the degree of the non-self-averaging effect for the maximal current and the diffusion coefficient.
\end{abstract}

\maketitle


\section{Introduction}
The one-dimensional asymmetric simple exclusion process (ASEP) is a pedagogical model for non-equilibrium systems \cite{Derrida}.
In particular, it describes various non-equilibrium phenomena such as traffic flow \cite{AritaFoulaadvandSanten} and protein synthesis 
by ribosomes \cite{ChouLakatou,CiandriniStansfieldRomano,DanaTuller}.
The ASEP is a stochastic process where particles with hard-core interactions diffuse on a one-dimensional lattice.
The ASEP can be mapped to a model of interface growth in the Kardar-Parisi-Zhang (KPZ) universality class \cite{KardarParisiZhang}.
Hopping to the right site in the ASEP corresponds to an increase in the interface.
The distribution of interface height was solved \cite{Johansson:2000tk,Tracy:2009tx,Aggarwal}.
Using the weak asymmetric limit of the ASEP, the KPZ equation was rigorously solved analytically \cite{PhysRevLett.104.230602,*SASAMOTO2010523,AmirGideon}.
Moreover, the large deviation function of the time-averaged current was obtained \cite{PhysRevLett.80.209,PhysRevLett.94.030601}.
The ASEP has been extended in various ways such as
Brownian ASEP~\cite{LipsRyabovMaass}, non-Poissonian hopping rates \cite{ConcannonBlythe},
and disordered hopping rates \cite{TripathyBarma,Enaud_2004,HarrisStinchcombe,JuhaszSantenIgloi,StinchcombeQueiroz,Nossan_2013,10.1214/14-BJPS277,BanerjeeBasu,PhysRevResearch.2.043073}.
When particles only hop to uni-direction, it is called the totally ASEP (TASEP).
For TASEPs, it is well known that the current-density relation is given by \cite{Derrida}
\begin{equation}
    J=\frac{1}{\tau}\rho(1-\rho),
    \label{J_TASEP}
\end{equation}
where $J$ is the particle current, $\rho$ is particle density, and $\tau$ is the inverse of the jump rate, i.e., the mean waiting time.
Moreover, in Refs.~\cite{Derrida_1993}, 
the variance of the tagged particle displacement, $\delta x_t$, in time $t$ is derived as a function of $\rho$:
\begin{equation}
    \frac{\braket{\delta x_t^2}-\braket{\delta x_t}^2}{t}\sim\frac{\sqrt{\pi}}{2\tau}\frac{(1-\rho)^{3/2}}{(L\rho)^{1/2}}
    \label{TASEP2}
\end{equation}
for $L\rightarrow\infty$ and $t\rightarrow\infty$, where $\braket{\cdot}$ is the ensemble average and $L$ is the system size.

Effects of disorder in the ASEP have been investigated for decades \cite{TripathyBarma,Enaud_2004,HarrisStinchcombe,JuhaszSantenIgloi,StinchcombeQueiroz,Nossan_2013,10.1214/14-BJPS277,BanerjeeBasu,PhysRevResearch.2.043073}.
Due to the disorder in the ASEP under the periodic boundary condition, 
a current-density relation deviates from that in the ASEP with a homogeneous jump rate, i.e., Eq.~(\ref{J_TASEP}).
More precisely, it becomes flat and the current is maximized on the flat regime \cite{TripathyBarma,HarrisStinchcombe,JuhaszSantenIgloi,StinchcombeQueiroz,Nossan_2013,10.1214/14-BJPS277,BanerjeeBasu}.
Moreover, in the flat regime, the low- and high-density phases coexist.
In the ASEP on networks, the flat regime also exists \cite{PhysRevLett.107.068702,Neri_2013,PhysRevE.92.052714}.
When the particle density is near $1/2$, the TASEP with short-ranged quenched disordered hopping rates
does not belong to the KPZ universality class but leads to a new universality class \cite{PhysRevResearch.2.043073}.
Under the open boundary condition, the first-order phase transition point between the low- and high-density phases
depends on the disorder \cite{Enaud_2004}.

Random walks in heterogeneous environments show anomalous diffusion.
The heterogeneous environment is characterized by a random energy landscape.
There are two types of random energy landscapes.
One is an annealed energy landscape, where the landscape randomly changes with time.
The continuous-time random walk is a diffusion model on the annealed energy landscape,
and its mean-squared displacement shows anomalous diffusion when the mean waiting time diverges \cite{METZLER20001}.
The other is a quenched energy landscape, where the landscape is configured randomly and does not change with time.
The quenched trap model (QTM) is a diffusion model on the quenched energy landscape \cite{BouchaudGeorfes}.
The mean-squared displacement of the QTM on an infinite system shows anomalous diffusion 
when the mean waiting time diverges \cite{BouchaudGeorfes}.
In the QTM on a finite system, the diffusion coefficient exhibits sample-to-sample fluctuations 
\cite{AkimotoBarkaiSaito,*AkimotoBarkaiSaito2018,LuoYi,AkimotoSaito2020}.
The diffusivity of interacting many-body systems on the annealed energy landscape
has been investigated \cite{Metzler:2014aa,Sanders_2014}.
However, the diffusivity of interacting many-body systems on the quenched energy landscape has never been investigated.
Such a heterogeneous environment is realized experimentally.
In protein synthesis by ribosomes, the codon decoding times become heterogeneous due to the heterogeneity of transfer RNA concentration \cite{DanaTuller}.
In other words, the distribution of the waiting time depends on the site, i.e., ribosomes diffuse on the quenched random environment.
There are other diffusion phenomena in such heterogeneous environments, such as train delays, 
proteins on DNA \cite{GraneliGreeneRobertsonYeykal,AustinCoxWang}, 
and water transportation in aquaporin \cite{AkimotoHiraoYamamotoYasuiYasuoka}.

In this paper, we investigate sample-to-sample fluctuations of the diffusivity for the TASEP on a quenched random energy landscape.
In our previous study, we show sample-to-sample fluctuations of the current \cite{https://doi.org/10.48550/arxiv.2208.10102}.
When an observable does not depend on the disorder realization, it is called self-averaging \cite{BouchaudGeorfes}.
In the QTM, it is known that the diffusion coefficient \cite{AkimotoBarkaiSaito,*AkimotoBarkaiSaito2018,LuoYi,AkimotoSaito2020},
the mobility \cite{AkimotoSaito2020}, and the mean first passage time \cite{AkimotoSaito2019} are non-self-averaging.
Is such a non-self-averaging behavior still observed when the $N$-body effect is introduced in the quenched random energy landscape?
This is a non-trivial question in diffusion in a heterogeneous environment.
In particular, it is non-trivial that the TASEP with disordered waiting-time distributions exhibits sample-to-sample fluctuations
for the current and the diffusion coefficient.
Therefore, it is important to provide an exact result for the current and the diffusion coefficient in heterogeneous quenched environments.

Our paper is organized as follows.
In Sec.~\ref{model}, we formulate the TASEP on a quenched random energy landscape and define averaging procedures.
In Sec.~\ref{numerical}, we show the numerical results of the current-density relation and the density profile.
In Sec.~\ref{derivation_density}, we present derivations of the density profile.
In Sec.~\ref{derivation_current}, we present derivations of the current and the diffusion coefficient.
In Sec.~\ref{SA}, we discuss the self-averaging properties of the current and the diffusion coefficient.
In Sec.~\ref{conclusion}, we conclude this paper.
In Appendix~\ref{a}, we derive the passage time distribution.
In Appendix~\ref{b}, we derive the Fr\'echet distribution.

\section{Model}\label{model}
We consider the TASEP on a quenched random energy landscape on a one-dimensional lattice.
It comprises $N$ particles on the lattice of $L$ sites with periodic boundary conditions.
Each site can hold at most one particle.
Quenched disorder means that when realizing the random energy landscape, it does not change with time.
At each lattice point, the depth $E>0$ of the energy trap is randomly assigned.
The depths are independent and identically distributed (IID) random variables with an exponential distribution, $\phi(E)=T_g^{-1}\exp{(-E/T_g)}$, 
where $T_g$ is called the glass temperature.
A particle can escape from a trap.
Escape times from a trap are IID random variables following an exponential distribution and
follow the Arrhenius law, i.e., the mean escape time of the $k$th site is given by $\tau_k=\tau_c\exp{(E_k/T)}$, 
where $E_k$ is the depth of the energy at site $k$, $T$ the temperature, and $\tau_c$ a typical time.
The probability of the escape time $\tau$ that is smaller than $x$ is given by $\Pr(\tau\leq x)\cong \Pr[E\leq T\ln(x/\tau_c)]$.
It follows that the probability density function (PDF) $\psi_\alpha(\tau)$ of waiting times follows a power-law distribution:
\begin{equation}
    \int_\tau^\infty d\tau'\psi_\alpha(\tau')\cong\left(\frac{\tau}{\tau_c}\right)^{-\alpha}\ (\tau\geq \tau_c)
    \label{eq1}
\end{equation}
with $\alpha\equiv T/T_g$ \cite{AkimotoBarkaiSaito,*AkimotoBarkaiSaito2018}. 

The dynamics of the particle are described by the Markovian one in the sense that the waiting time is memory-less. In particular, 
the waiting times at site $k$ are assigned IID random variables following an exponential distribution,
$\psi_k(t_i)=\tau_k^{-1}\exp{(-t_i/\tau_k)}$.
After the waiting time elapses, the particle attempts to hop the neighboring site on its right. The hop is accepted only if the site is empty.
When the attempt is a success or failure, the particle is assigned a new waiting time from $\psi_{k+1}(t_i)$ or $\psi_k(t_i)$, respectively.

Here, we consider three averaging procedures, i.e., ensemble average, disorder average, and time average.
The ensemble average of observable $\mathcal{O}(t)$ is an average with respect to a stationary ensemble for a single disorder realization denoted by $\langle\mathcal{O}(t)\rangle$.
The disorder average of observable $\mathcal{O}(t)$ is an average with respect to different disorder realizations denoted by $\langle\mathcal{O}(t)\rangle_{\mathrm{dis}}$.
The time average of observable $\mathcal{O}(t)$ is defined by
\begin{equation}
    \bar{\mathcal{O}}(T)=\frac{1}{T}\int_0^T\mathcal{O}(t)dt.
    \label{time_average}
\end{equation}
Furthermore, we consider a stationary initial condition.
For the ASEP on a finite system, the variance of the displacement of the tagged particle depends on whether the initial conditions are
identical or not, especially for a short time \cite{GuptaMajumdarGodrecheBarma}.
However, the asymptotic behavior does not depend on the initial condition.
In this paper, we are interested in the asymptotic behavior of the current and the diffusivity.
Therefore, the initial conditions in this paper are not fixed.
In numerical simulations, particles start from the stationary ensemble of configurations.
The stationary ensemble is given by the configuration after particles arrange randomly and diffuse for a long time.

\section{Numerical results of current-density relation and density profile}\label{numerical}
\begin{figure}[b]
    \centering
    \includegraphics[width=8.6cm]{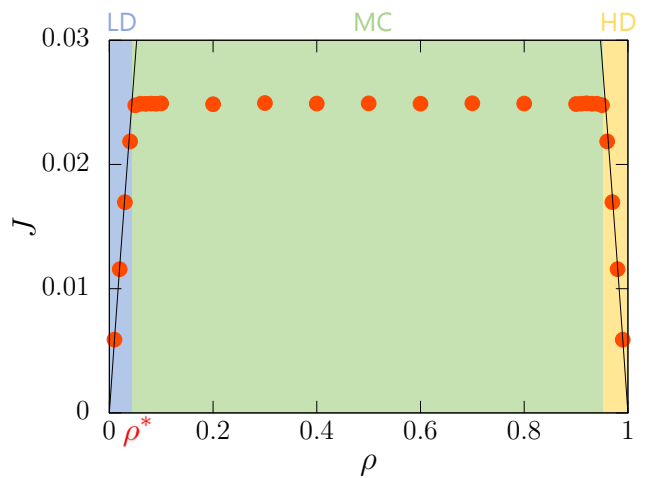}
    \caption{Current-density relations for homogeneous and disordered TASEPs. 
    The circles are obtained by the numerical simulation of dynamics of the DTASEP 
    ($L=5000$, $\alpha=2.5$, and $\tau_c=1$). The solid line represents the current-density relation, Eq.~(\ref{J_TASEP}),
    for the homogeneous TASEP with $\tau$ being set to equal to the sample average of the waiting times of the DTASEP.
    $\rho^*$ is given by Eq.~(\ref{rho_max}).}
    \label{fig:TASEP_J}
\end{figure}

We numerically show that the current-density relation
for a disordered TASEP (DTASEP) deviates from that for a TASEP with a homogeneous jump rate, i.e., the homogeneous TASEP.
Figure~\ref{fig:TASEP_J} shows the steady-state current $J$ against particle density $\rho=N/L$, i.e., the current-density relation,
for a DTASEP. 
For low and high densities, the current-density relation is the same as that of the homogeneous TASEP (see Fig.~\ref{fig:TASEP_J}).
However, there is a distinct difference between them in the intermediate regime. 
In particular, the current for the DTASEP becomes almost 
flat and smaller than that for the homogeneous TASEP in the intermediate regime.
On the other hand, there is no flat regime for the homogeneous TASEP. 
The flat regime in the DTASEP is observed in other disordered systems \cite{TripathyBarma,HarrisStinchcombe,JuhaszSantenIgloi,StinchcombeQueiroz,BanerjeeBasu}.
Thus, it is a manifestation of the existence of a disorder. 
In this regime, the current is independent of the particle density and maximized.
In the following, we classify the density into three regimes: the low density (LD) ($0<\rho\leq \rho^*$),
the maximal current (MC) ($\rho^*<\rho<1-\rho^*$), and the high density (HD) ($1-\rho^*\leq \rho<1$) regimes (Fig.~\ref{fig:TASEP_J}).
We explicitly derive the transition density $\rho^*$ later (see Eq.~(\ref{rho_max})).

\begin{figure*}
    \centering
    \subfigure[\hspace{6cm}]{\includegraphics[width = 7cm]{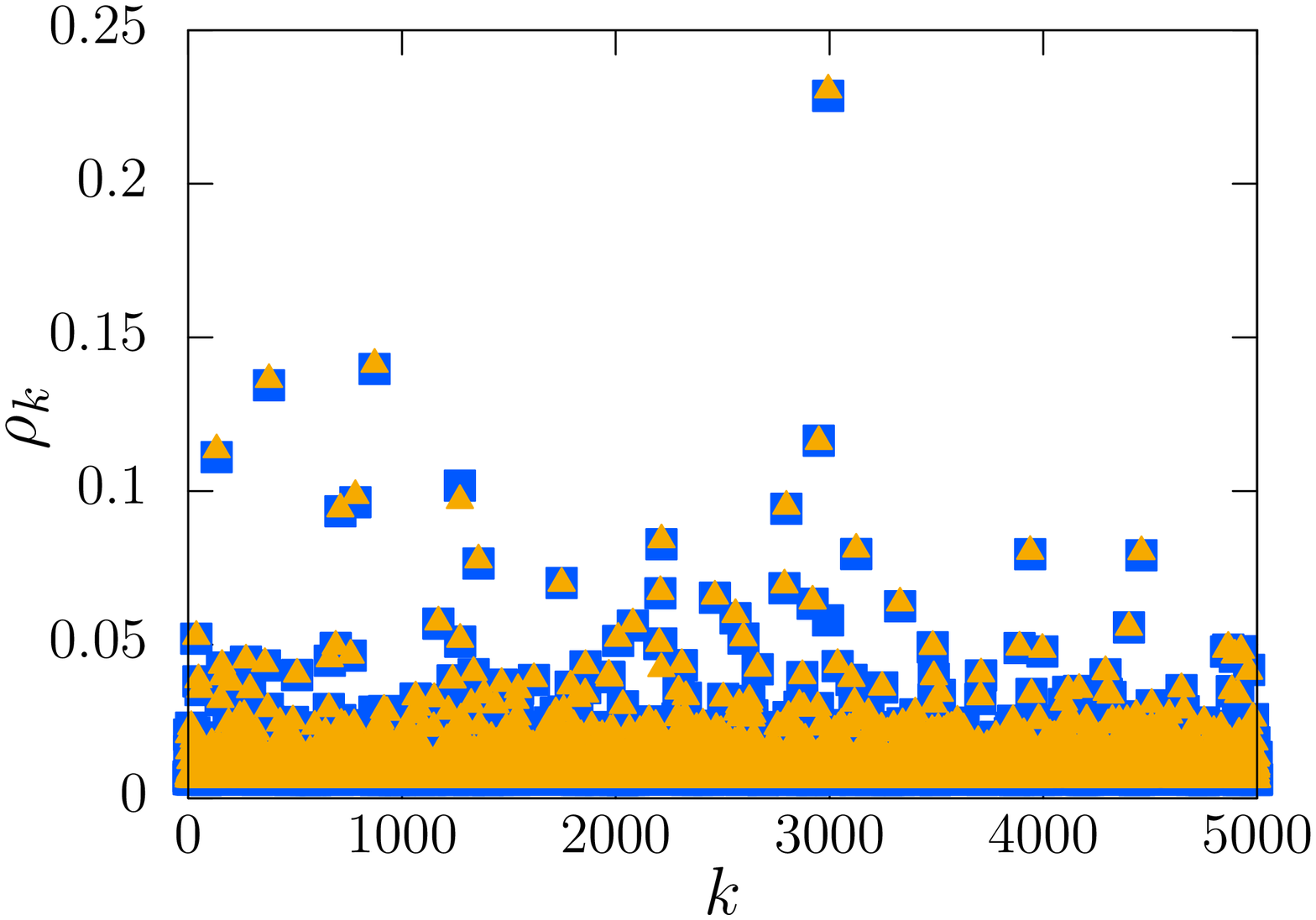}\label{fig:rho_50}}
    \subfigure[\hspace{6cm}]{\includegraphics[width = 7cm]{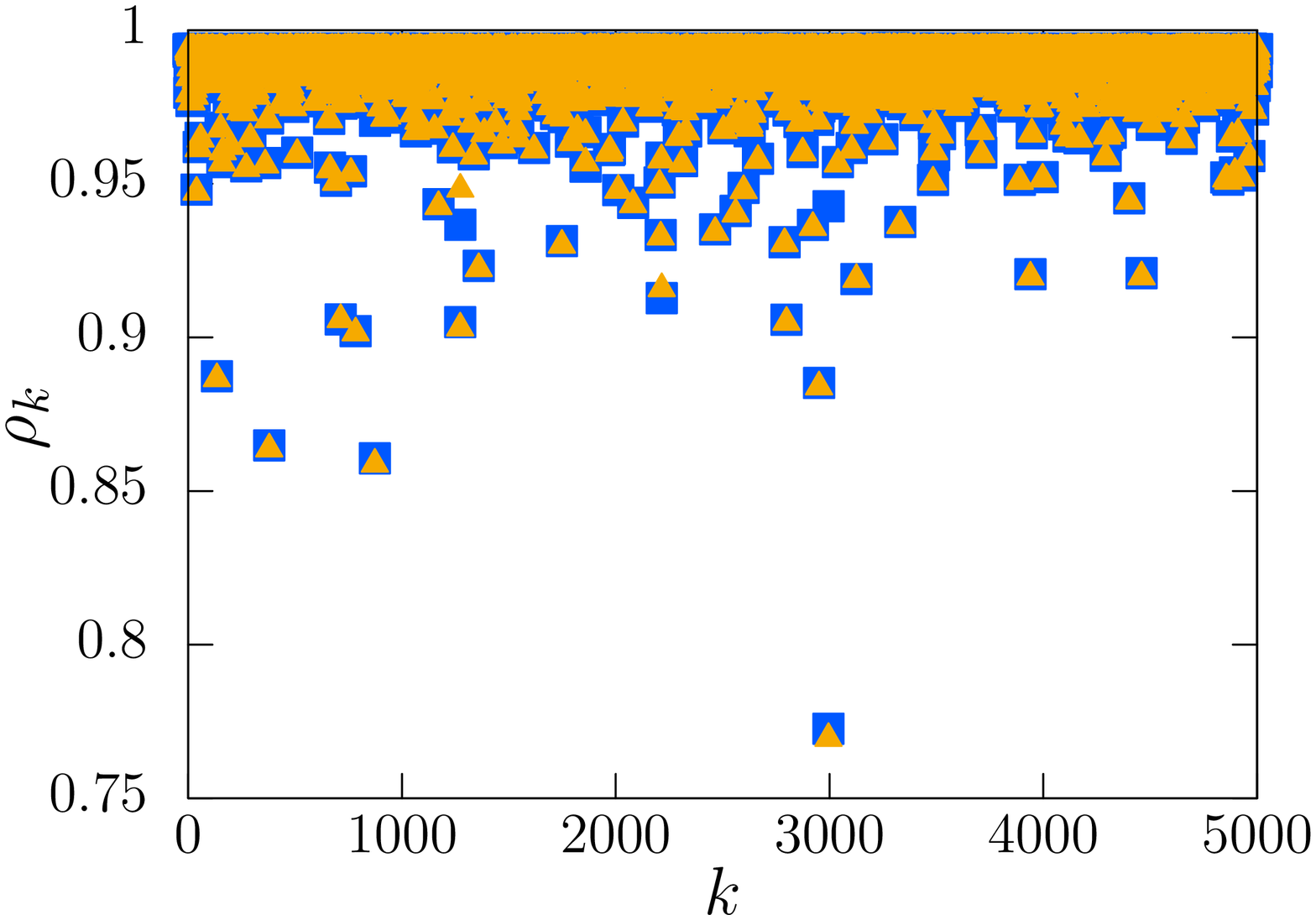}\label{fig:rho_4950}}\\
    \subfigure[\hspace{6cm}]{\includegraphics[width = 7cm]{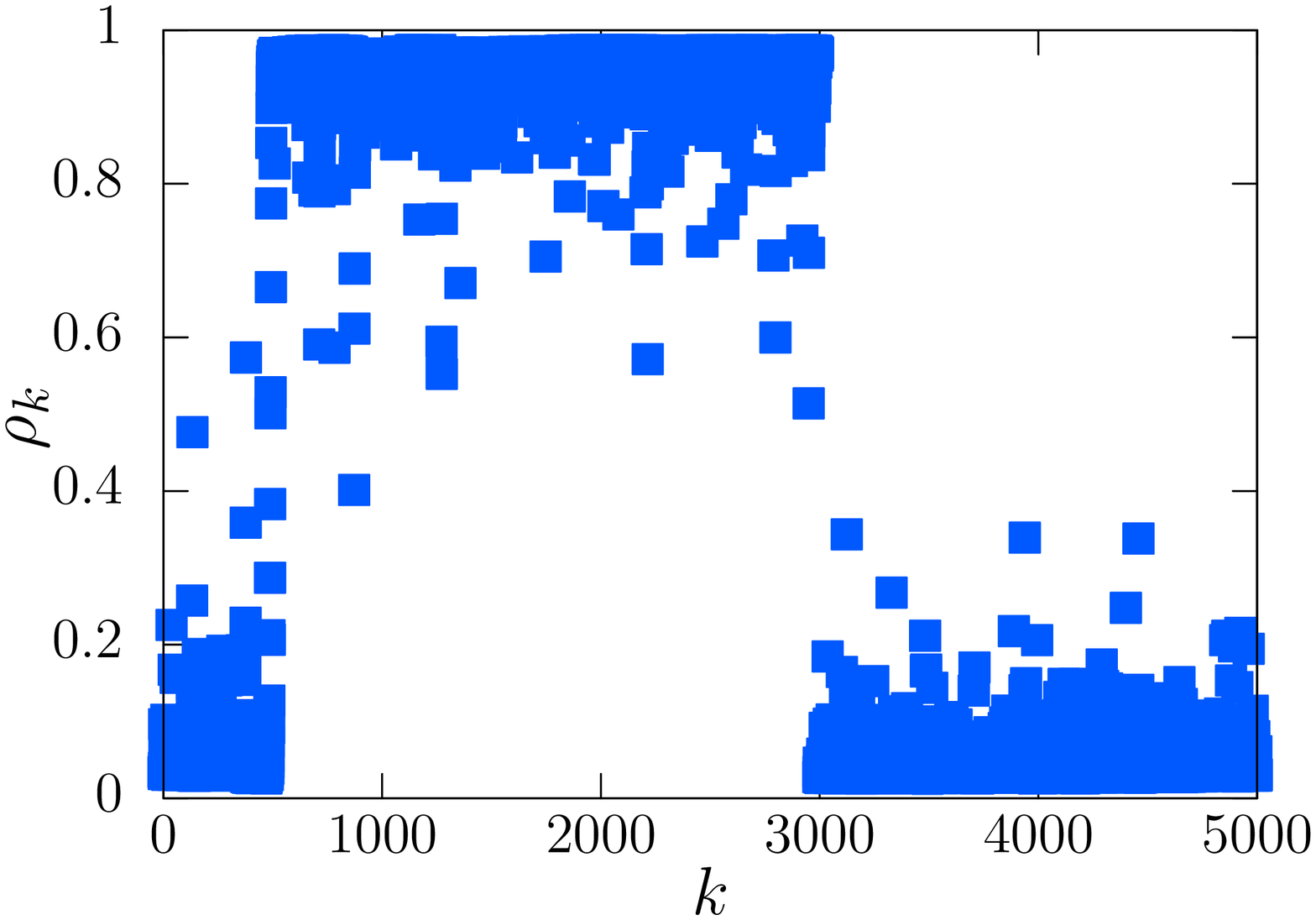}\label{fig:rho_2500}}
    \subfigure[\hspace{6cm}]{\includegraphics[width = 7cm]{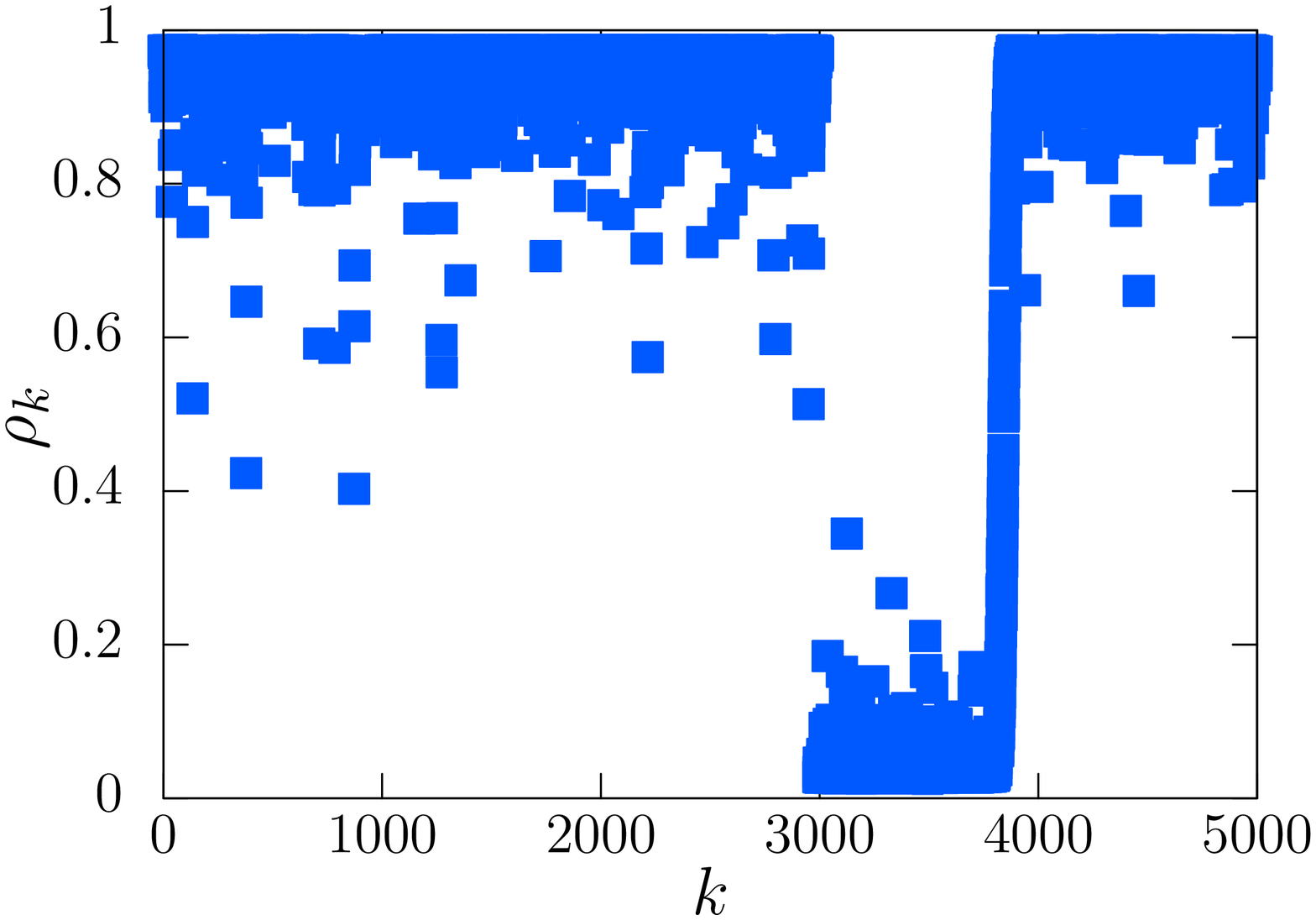}\label{fig:rho_4000}}
    \caption{Density profiles: (a) $\rho=0.01$, (b) $\rho=0.99$, (c) $\rho=0.5$, and (d) $\rho=0.8$ ($L=5000$, $\alpha=2.5$, and $\tau_c=1$). The squares are the results of the numerical simulation of the dynamics of the DTASEP. Triangles are Eqs. (\ref{rho_LD}) and (\ref{rho_HD}) for (a) and (b), respectively.}
    \label{fig:QTM_rho}
\end{figure*}

Here, we numerically show the density profiles.
For the LD and HD regimes, the system is homogeneous on a macroscopic scale (Figs.~\ref{fig:QTM_rho}\subref{fig:rho_50} and \subref{fig:rho_4950}).
For the MC regime, there is a macroscopic density segregation (Figs.~\ref{fig:QTM_rho}\subref{fig:rho_2500} and \subref{fig:rho_4000}).
The segregation is classified into high- and low-density phases by the deepest trap.
Comparing Figs.~\ref{fig:QTM_rho}\subref{fig:rho_2500} and \subref{fig:rho_4000}, we observe that the high-density phase becomes large when the particle density is increased.
This result is qualitatively similar to that in a system with one defect bond, studied in Ref.~\cite{JanowskyLebowitz}.

We discuss the properties of the current-density relation in the DTASEP,
in particular, why the maximal current does not depend on the particle density $\rho$.
The phase separation in the density profile occurs because of a traffic jam caused by the site of the maximum mean waiting time.
The local particle densities in the LD and HD phases become constant (see Sec.~\ref{derivation_density}).
As a result, the maximum current becomes constant, i.e., it does not depend on the particle density $\rho$.
Furthermore, since the phase separation in the density profile does not occur suddenly,
the transition from the LD regime to the MC regime must be continuous.

\section{Derivation of the density profile}\label{derivation_density}
Here, we derive the density profile by the mean-field approximation.
This derivation is almost the same as our previous study \cite{https://doi.org/10.48550/arxiv.2208.10102}.
Let $J_k$ be the mean current across the bond between site $k$ and $k+1$. In the DTASEP, a hop occurs with a rate $1/\tau_k$ whenever site $k$ is occupied, and site $k+1$ is not. Thus, the mean current is represented by
\begin{equation}
    J_k=\Braket{\frac{1}{\tau_k}n_k(1-n_{k+1})},
    \label{J_k}
\end{equation}
where $n_k$ denotes the number of a particle, which is $1$ if the site $k$ is occupied and $0$ otherwise. 
If the system is in a steady state, the ensemble average is equal to the time average in the long-time limit, i.e., the system is ergodic.
The ensemble average in Eq. (\ref{J_k}) coincides with the long-time average
if the system is ergodic.
The periodic boundary condition implies $n_{L+1}=n_1$ and $\tau_{L+1}=\tau_1$. 
The probability of finding a particle at site $k$ is given by $\rho_k=\braket{n_k}$.
In the mean-field approximation, one can ignore correlations between $n_k$ and $n_{k+1}$, which means
\begin{equation}
    \braket{n_kn_{k+1}}=\braket{n_k}\braket{n_{k+1}}.
\end{equation} 
In the steady state, the site densities are time-independent. 
Moreover, from the continuity of the current, the current is independent of $k$, i.e., $J_k=J$ for all $k$. Therefore, we have 
the current-density relation:
\begin{equation}
    J=\frac{1}{\tau_k}\rho_k(1-\rho_{k+1}).
    \label{J}
\end{equation}
We note that the right-hand side of Eq.~(\ref{J}) is independent of $k$. 

We derive a simpler form of the site density by approximating Eq.~(\ref{J}) for the LD and HD regimes.
For the LD regime,  we can assume $\rho_k\rho_{k+1}\ll 1$ because the particle density is small.
Ignoring $\rho_k\rho_{k+1}$ in Eq.~(\ref{J}), we obtain
\begin{equation}
    J\cong \frac{1}{\tau_k}\rho_k.
    \label{J_app}
\end{equation}
Using the conservation of particles, $\sum_i\rho_i=N$, the site density has the form
\begin{equation}
    \rho_k\cong\frac{\tau_k}{\mu}\rho,
    \label{rho_LD}
\end{equation}
for the LD regime, where $\mu$ is the sample average of the waiting times, $\mu=\sum_i\tau_i/L$.
This result is the same as the steady-state density for the QTM \cite{AkimotoBarkaiSaito}. 
For the HD regime, the particle density is high.
Using the hole density, $\sigma_k=1-\rho_k$, instead of $\rho_k$, we can derive the site density in the same way as 
in the LD regime. The result becomes 
\begin{equation}
    \rho_k=1-\sigma_k\cong1-\frac{\tau_{k-1}}{\mu}(1-\rho).
    \label{rho_HD}
\end{equation}
Figures \ref{fig:QTM_rho}\subref{fig:rho_50} and \ref{fig:QTM_rho}\subref{fig:rho_4950} show the density profiles for LD and HD regimes, respectively.
The densities are well described by the set of site densities $\{\rho_k\}$.
Therefore, Eqs. (\ref{rho_LD}) and (\ref{rho_HD}) are good approximated forms of the site densities. 
The results for the LD and HD regimes reproduce the current-density relation for a homogeneous TASEP. 
In other words, the system is homogeneous on a macroscopic scale. 

Next, we approximately obtain a density $\rho^*$ which is the boundary density between LD and MC regimes in the current-density relation 
(see Fig.~\ref{fig:TASEP_J}).
The current in the MC regime does not depend on the density $\rho$,
and let $J_{\max}$ be the maximal current.
We define the boundary density $\rho^*$ by the point at which $J_{\max}$ is equal to $\mu^{-1}\rho(1-\rho)$,
\begin{equation}
    J_{\max}=\frac{1}{\mu}\rho^*(1-\rho^*).
\end{equation}
Solving this equation for $\rho^*$, we have
\begin{equation}
    \rho^*=\frac{1-\sqrt{1-4\mu J_{\max}}}{2}.
\end{equation}
For the large-$L$ limit, $J_{\max}$ is much smaller than the maximal current for the homogeneous TASEP, i.e.,
$J_{\max}\ll 1/(4\mu)$.
Hence, we can approximate the boundary density by
\begin{equation}
    \rho^*\sim\frac{1}{2}-\frac{1}{2}(1-2\mu J_{\max})=\mu J_{\max}.
    \label{rho_MC}
\end{equation}

We derive the site density in the MC regime.
The current in the MC regime does not depend on the site, i.e., Eq.~(\ref{J}) is valid,
\begin{equation}
    J_{\max}=\frac{1}{\tau_k}\rho_k(1-\rho_{k+1}).
    \label{Jmax_rela}
\end{equation}
Using Eq.~(\ref{rho_LD}), the site density in the LD phase is given by $\rho_k\cong\tau_k\rho_{LD}/\mu$,
where $\rho_{LD}$ is the particle density in the LD phase.
When both sites $k$ and $k+1$ exist in the LD phase, we can ignore $\rho_k\rho_{k+1}$ due to the low-density limit.
Therefore, the maximal current is given by $J_{\max}\sim\rho_{LD}/\mu$.
Furthermore, the particle density in the LD phase becomes $\rho_{LD}\sim\rho^*$.
Thus, the site density in the LD phase is represented by
\begin{equation}
    \rho_k\sim\frac{\tau_k}{\mu}\rho^*.
    \label{rho_MC_LD}
\end{equation}
We can also derive the site density in the HD phase in the same way as in the LD phase.
The site density in the HD phase is represented by
\begin{equation}
    \rho_k\sim 1-\frac{\tau_{k-1}}{\mu}\rho^*.
    \label{rho_MC_HD}
\end{equation}

We derive the maximal current based on the phase separation of the density profile in the MC regime
(Figs.~\ref{fig:QTM_rho}\subref{fig:rho_2500} and \subref{fig:rho_4000}).
We numerically find that the site with the maximal mean waiting time is always the boundary between the HD and the LD phases.
When the mean waiting time is maximized at site $m$, sites $m$ and $m+1$ exist in high- and low-density phases, respectively.
The site densities at site $m$ and $m+1$ are given by Eqs.~(\ref{rho_MC_HD}) and (\ref{rho_MC_LD}), respectively, i.e.,
$\rho_m\sim 1-\tau_{m-1}\rho^*/\mu$ and $\rho_{m+1}\sim\tau_{m+1}\rho^*/\mu$.
Using these values and Eq.~(\ref{rho_MC}), Eq.~(\ref{Jmax_rela}) is represented by
\begin{equation}
    \begin{split}
        J_{\max}&=\frac{1}{\tau_m}\rho_m(1-\rho_{m+1})\\
        &\sim\frac{1}{\tau_m}(1-\tau_{m-1}J_{\max})(1-\tau_{m+1}J_{\max}).
    \end{split}
\end{equation}
Ignoring the quadratic term of $J_{\max}$ and solving this equation,
we obtain the maximal current
\begin{equation}
    J_{\max}\sim\frac{1}{\tau_{m-1}+\tau_m+\tau_{m+1}}.
\end{equation}
In the following, we assume that the mean waiting time is maximized at site $m$.
For $L\rightarrow\infty$, $\tau_m$ is much longer than $\tau_{m-1}$ and $\tau_{m+1}$, i.e., $J_{\max}\sim\tau_m^{-1}$.
Therefore, we obtain the boundary density
\begin{equation}
    \rho^*\sim\frac{\mu}{\tau_m}.
    \label{rho_max}
\end{equation}
By the extreme value theory \cite{HaanFerreira}, the scaling of $\tau_m$ follows
\begin{equation}
    \tau_m=O(L^{1/\alpha})
\end{equation}
for $L\rightarrow\infty$.
For $\alpha>1$, the first moment of the waiting times exists; i.e., 
$\mu\rightarrow\braket{\tau}\equiv \int_0^\infty\tau\psi_\alpha(\tau)d\tau$ ($L\rightarrow\infty$).
Hence, the scaling of $\rho^*$ becomes
\begin{equation}
    \rho^*\propto L^{-1/\alpha}.
    \label{rho*}
\end{equation}
For $\alpha\leq 1$, the first moment of the waiting times diverges.
The scaling of the sum of $\tau_i$ follows
\begin{equation}
    \sum_{i=1}^L\tau_i=O(L^{1/\alpha})
\end{equation}
for $L\rightarrow\infty$.
It follows that the scaling of $\rho^*$ becomes
\begin{equation}
    \rho^*\sim L^{-1}\frac{\sum_i\tau_i}{\tau_m}\propto L^{-1}
    \label{rho^*_L}.
\end{equation}
Therefore, $\rho^*\rightarrow 0$ for $L\rightarrow\infty$.

We derive the location of the shock.
Since the HD phase occurs due to the site with the maximum mean-waiting time,
we consider the distance from the site with the maximum mean-waiting time to the shock, i.e., the length of the HD phase $l_h$.
The local particle densities in the LD and HD phases are given by $\rho^*$ and $1-\rho^*$, respectively.
Based on the conservation of particles, the number of particles is represented by
\begin{equation}
    L\rho=l_h(1-\rho^*)+(L-l_h)\rho^*.
\end{equation}
Solving this equation for $l_h$, we have
\begin{equation}
    l_h=\frac{L(\rho-\rho^*)}{1-2\rho^*}.
\end{equation}
Therefore, the length of the HD phase increases with the density, and that is consistent with the numerical results 
(Figs.~\ref{fig:QTM_rho}\subref{fig:rho_2500} and \subref{fig:rho_4000}).

\begin{figure*}
    \centering
    \subfigure[\hspace{4cm}]{\includegraphics[width = 4.7cm]{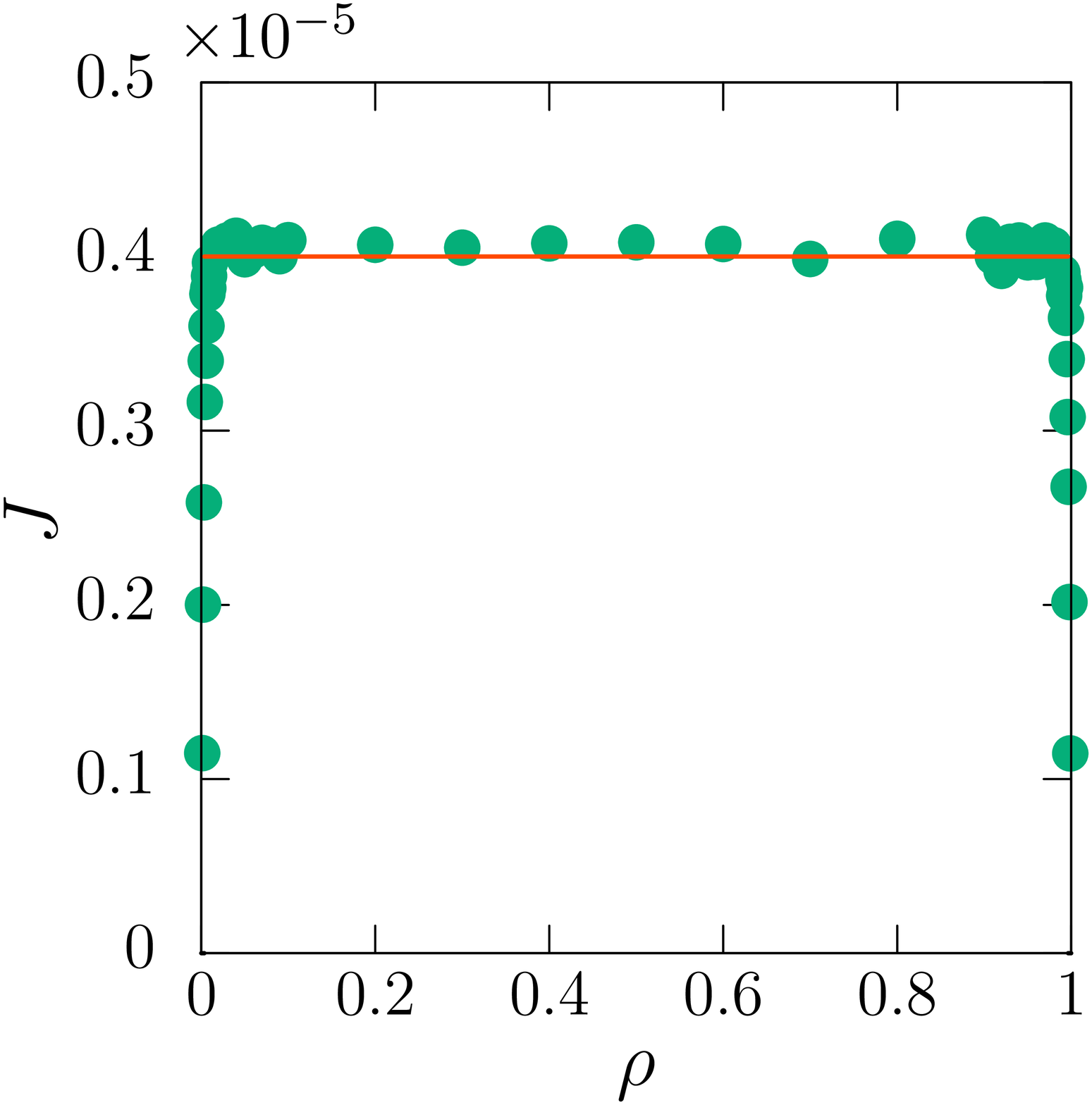}\label{fig:J_0.5}}
    \subfigure[\hspace{4cm}]{\includegraphics[width = 5cm]{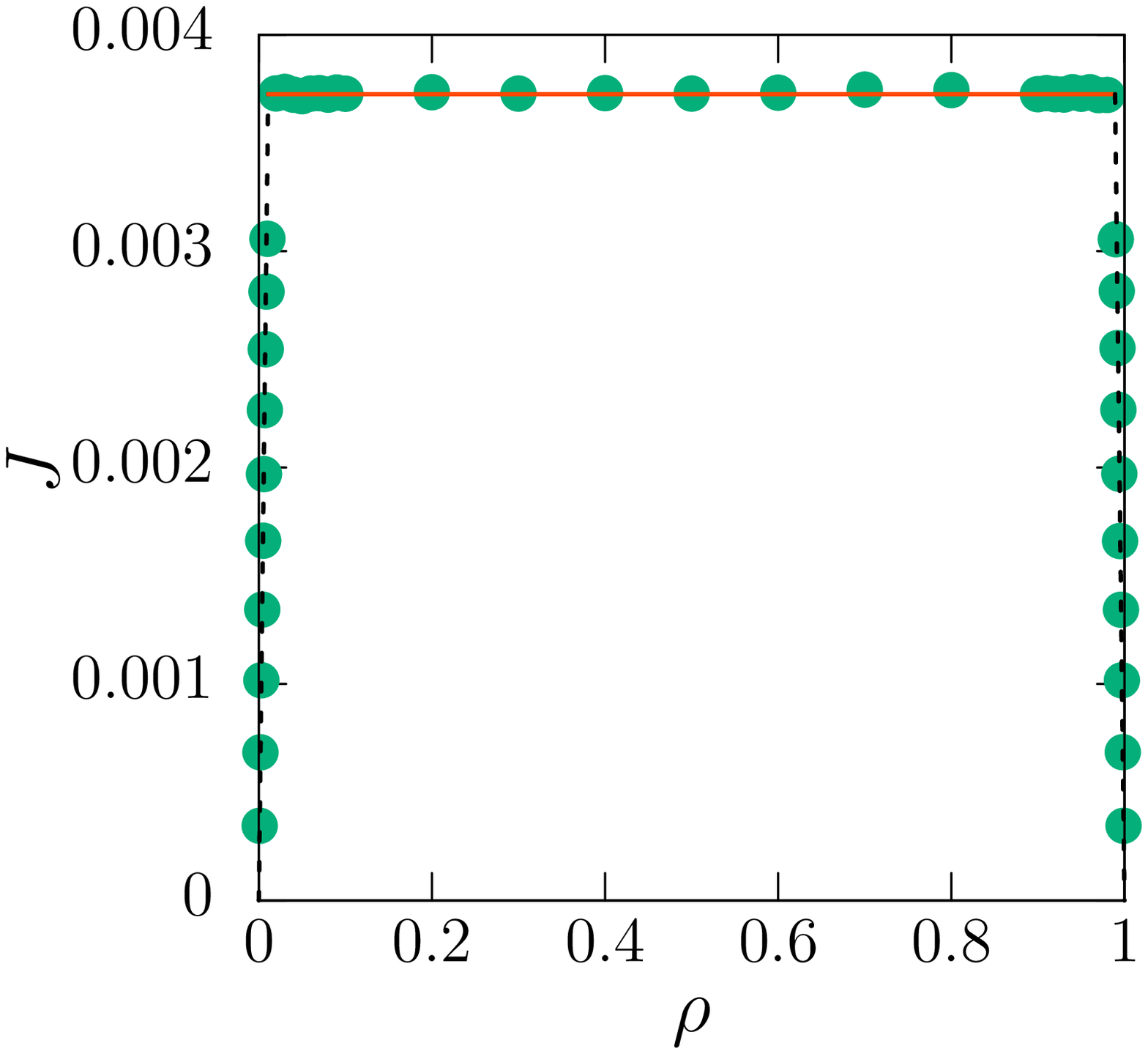}\label{fig:J_1.5}}
    \subfigure[\hspace{4cm}]{\includegraphics[width=5cm]{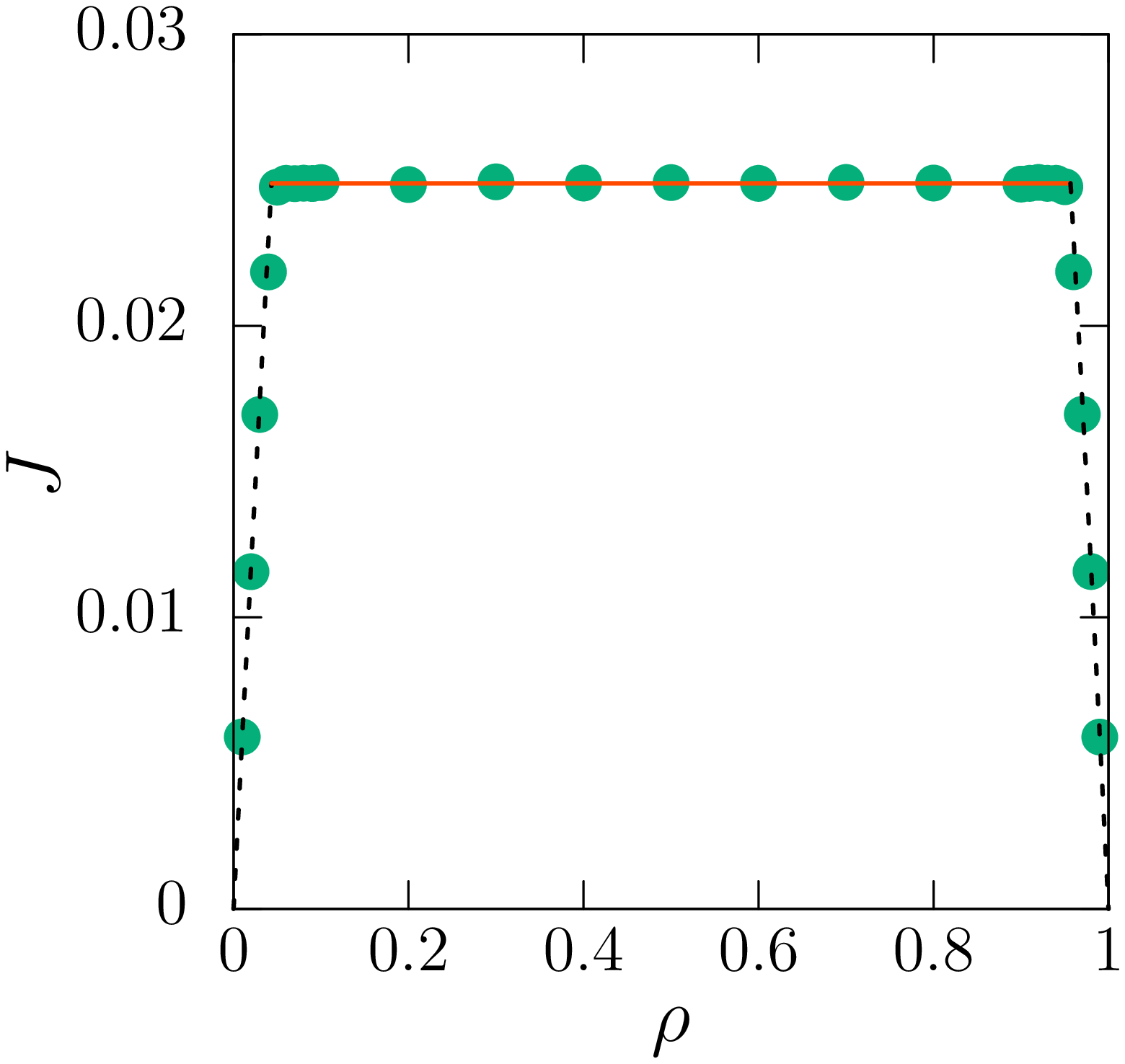}\label{fig:J_2.5}}
    \caption{Current-density relation for different $\alpha$, i.e.,
    (a) $\alpha=0.5$, (b) $\alpha=1.5$, and (c) $\alpha=2.5$, where the disorder realizations are fixed.
    The circles are obtained by the numerical simulation of the dynamics of the DTASEP ($L=1000$ for (a) and $5000$ for other cases). 
    The dashed and the solid lines represent Eqs.~(\ref{J_LD}) and (\ref{J_max}), respectively.}
    \label{fig:QTM_J}
\end{figure*}

\begin{figure*}
    \centering
    \subfigure[\hspace{4cm}]{\includegraphics[width = 5cm]{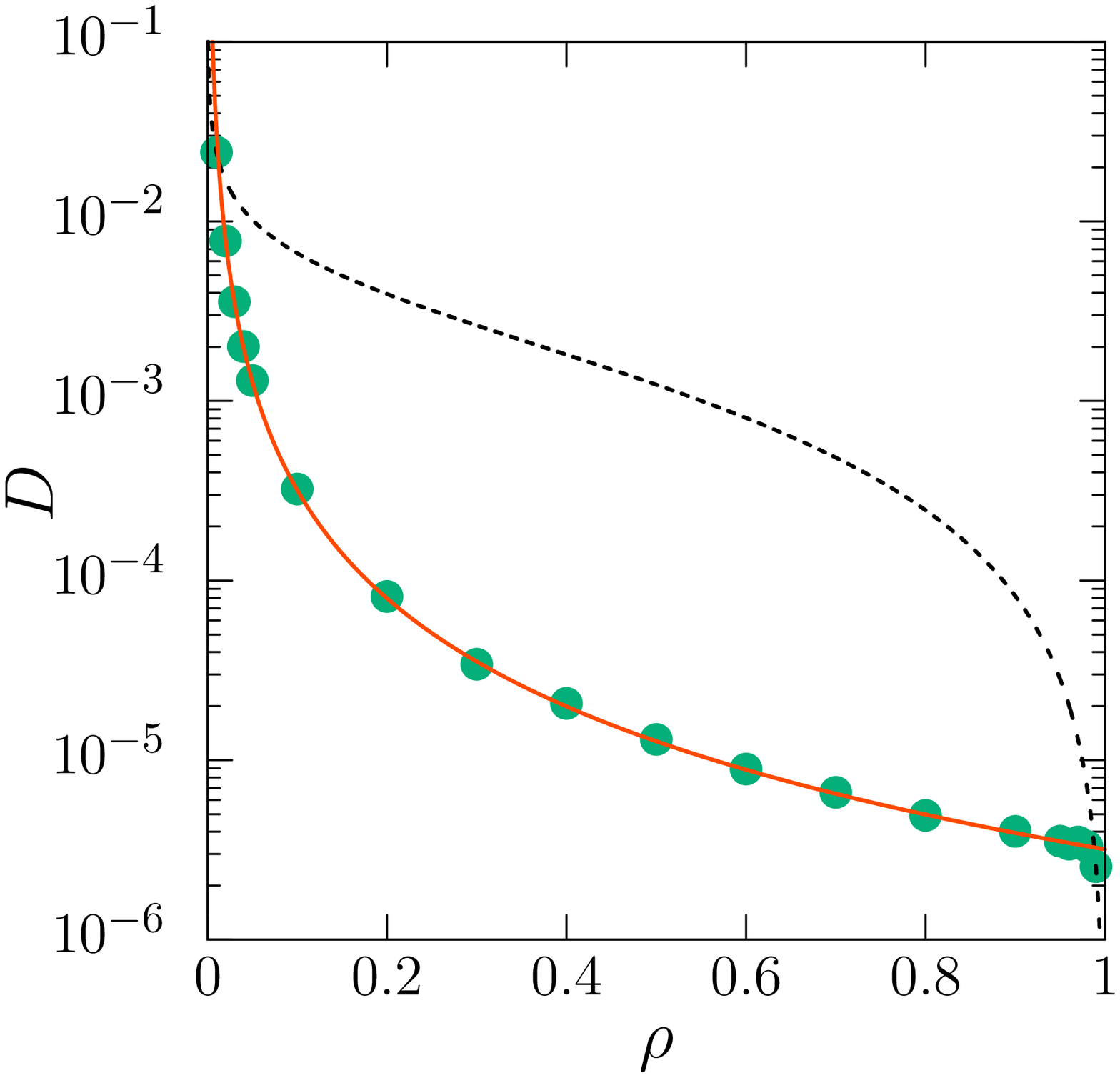}\label{fig:D_0.5}}
    \subfigure[\hspace{4cm}]{\includegraphics[width = 5cm]{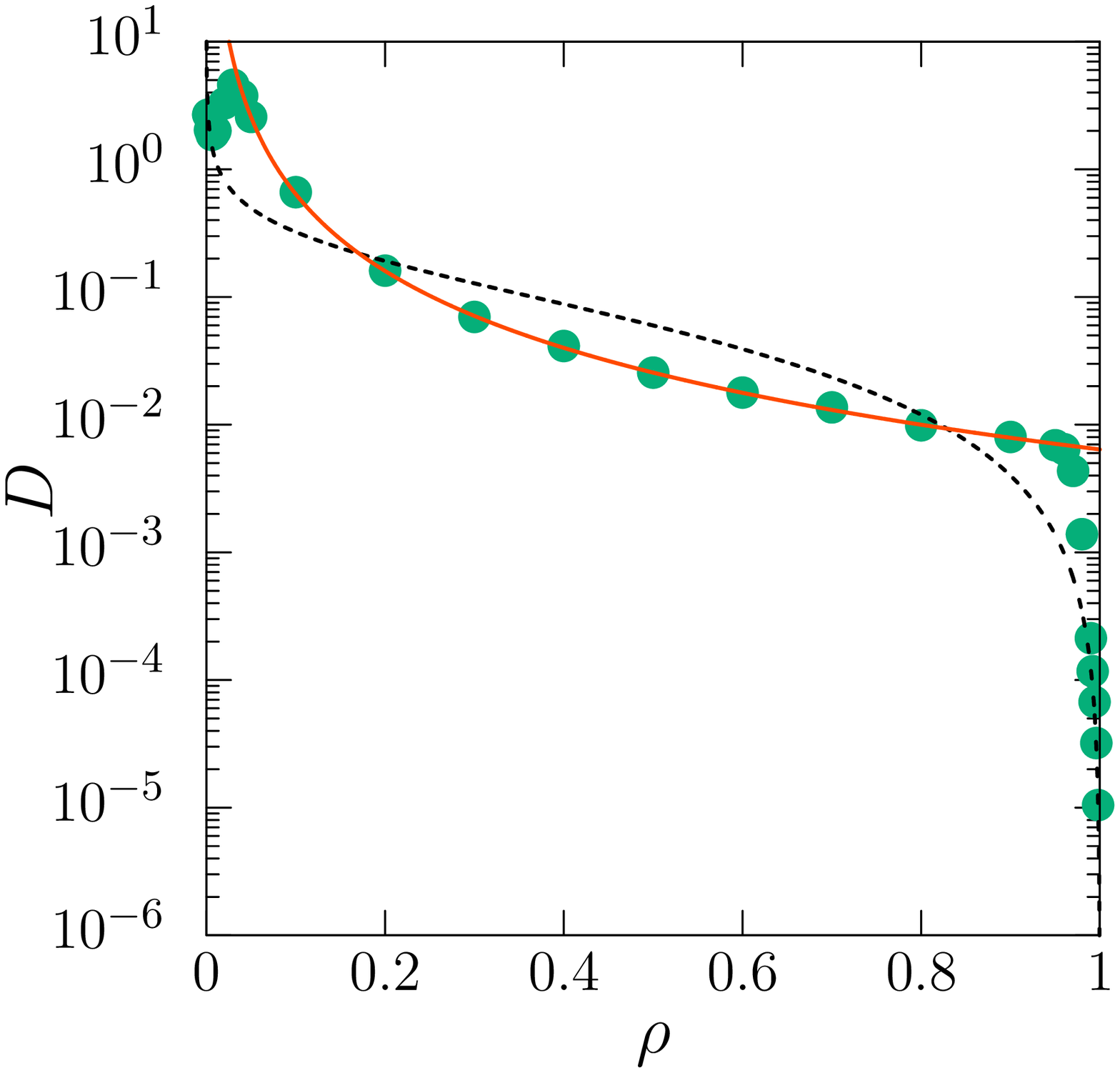}\label{fig:D_1.5}}
    \subfigure[\hspace{4cm}]{\includegraphics[width=5cm]{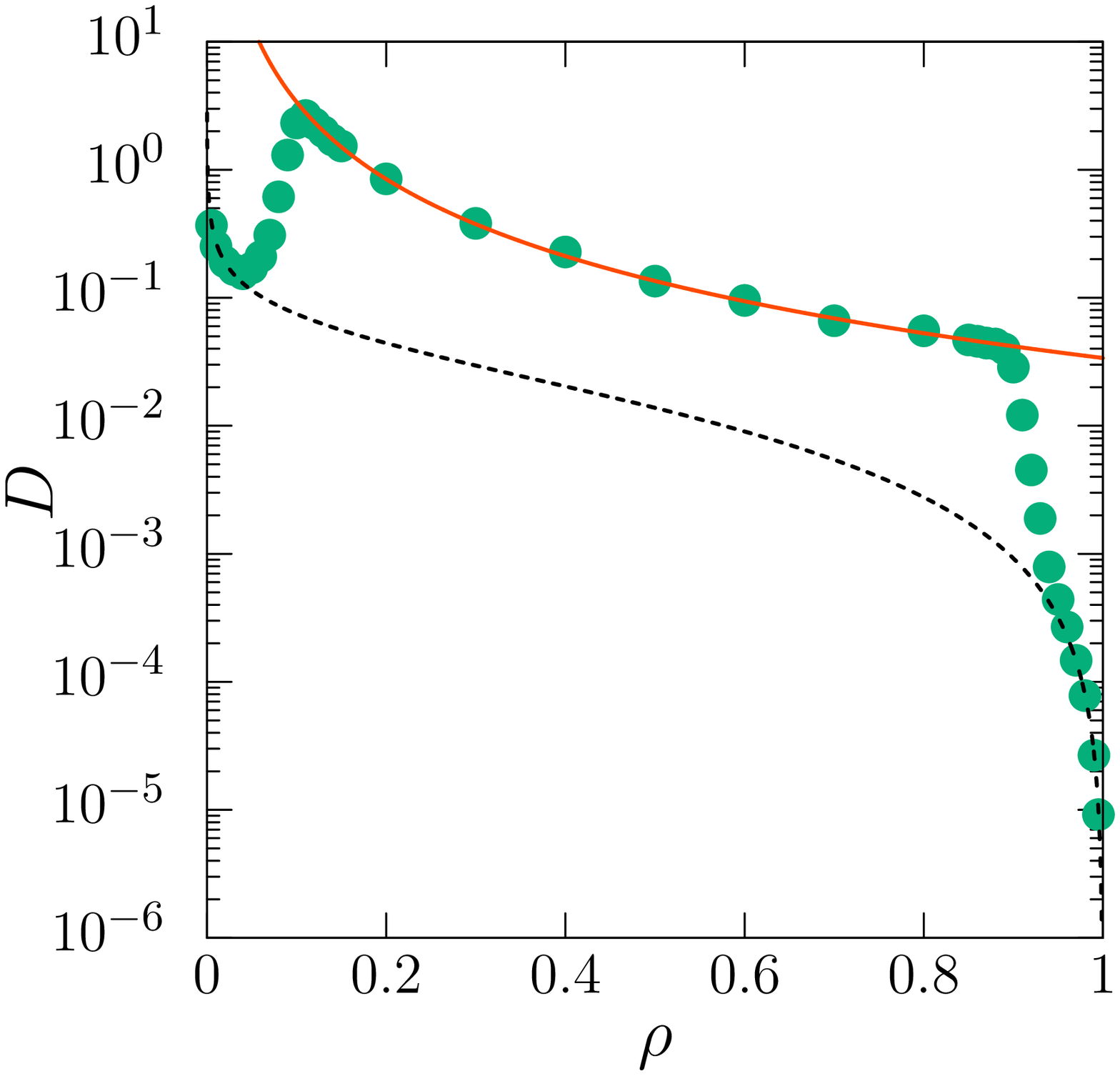}\label{fig:D_2.5}}
    \caption{Diffusion coefficient-density relation for different $\alpha$, i.e.,
    (a) $\alpha=0.5$, (b) $\alpha=1.5$, and (c) $\alpha=2.5$, where the disorder realizations are fixed.
    The circles are obtained by the numerical simulation of the dynamics of the DTASEP ($L=100$ for (a), $500$ for (b), and $1000$ for (c)).
    The dashed and the solid lines represent Eqs.~(\ref{D_LD}) and (\ref{D_MC}), respectively.}
    \label{fig:QTM_D}
\end{figure*}
\section{Derivation of current and diffusivity}\label{derivation_current}
\subsection{LD and HD regimes}
Here, we derive the current in the LD and HD regimes.
For single-particle dynamics on the quenched random energy landscape, i.e., the QTM, the mean number of events
that a particle passes a site until time $t$ is given by \cite{AkimotoSaito2020}
\begin{align}
    \frac{\braket{Q_t}}{t}\sim\frac{1}{L\mu}\ \ (t\rightarrow\infty),\label{QTM1}
\end{align}
where $Q_t$ is the number of events that a particle passes a site until time $t$.
For the DTASEP in the LD and HD regimes, the current depends on the particle density,
which is identical for the homogeneous TASEP (Eq.~(\ref{J_TASEP})).
Hence, the current in the LD and HD regimes is given by
\begin{equation}
    J\sim a\rho(1-\rho)
\end{equation}
for $L\rightarrow\infty$.
When $\rho=1/L$, the current is equal to Eq.~(\ref{QTM1}) for $L\rightarrow\infty$, i.e.,
the constant $a$ is given by $a=1/\mu$.
Therefore, we have the current in the LD and HD regimes:
\begin{equation}
    J\sim\frac{1}{\mu}\rho(1-\rho)\label{J_LD}
\end{equation}
for $L\rightarrow\infty$.

Next, we derive the diffusion coefficient in the LD and HD regimes.
$\delta x_t$ denotes the displacement of the tagged particle until time $t$.
For the QTM, the variance of the displacement is given by \cite{AkimotoSaito2020}
\begin{equation}
    \lim_{t\rightarrow\infty}\frac{\braket{\delta x_t^2}-\braket{\delta x_t}^2}{t}\sim \frac{\sigma^2}{\mu^3}\label{QTM2}
\end{equation}
for $L\rightarrow\infty$, where $\sigma^2$ is the sample mean of the squared waiting times, $\sigma^2=\sum_i\tau_i^2/L$.
For the DTASEP in the LD and HD regimes, the variance of the displacement depends on the particle density,
which is identical for the homogeneous TASEP (Eq.~(\ref{TASEP2})).
Hence, the diffusion coefficient, $D\equiv\lim_{t\rightarrow\infty}(\braket{\delta x_t^2}-\braket{\delta x_t}^2)/t$, is given by
\begin{equation}
    D\sim b\frac{\sqrt{\pi}}{2}\frac{(1-\rho)^{3/2}}{\rho^{1/2}}L^{-1/2}
\end{equation}
for $L\rightarrow\infty$.
When $\rho=1/L$, the diffusion coefficient is equal to Eq.~(\ref{QTM2}) for $L\rightarrow\infty$, i.e.,
the constant $b$ is given by $b=2\sigma^2/\mu^3\sqrt{\pi}$.
The diffusion coefficient in the LD and HD regimes is given by
\begin{align}
    D\sim\frac{\sigma^2}{\mu^3}\frac{(1-\rho)^{3/2}}{\rho^{1/2}}L^{-1/2}\label{D_LD}
\end{align}
for $L\rightarrow\infty$.

\subsection{MC regime}
Here, we derive the maximal current and the diffusion coefficient in the MC regime by the renewal theory.
We define the passage time as a time interval between consecutive events that particles pass a site.
We note that the passage time differs from the first passage time because the particles which pass a site are different.
When the target site is $m$, the mean and the variance of the passage time $T_m$ are obtained in Ref.~\cite{https://doi.org/10.48550/arxiv.2208.10102}
(see also Appendix~\ref{a}):
\begin{equation}
    \braket{T_m}=\tau_m+\frac{\tau_{m-1}}{\rho_{m-1}}+\frac{\frac{\rho_{m-1}}{\tau_{m-1}}}{\frac{\rho_{m-1}}{\tau_{m-1}}+\frac{1-\rho_{m+2}}{\tau_{m+1}}}\frac{\tau_{m+1}}{1-\rho_{m+2}},
    \label{tau_ave}
\end{equation}
\begin{equation}    
    \begin{split}
        \braket{T_m^2}-\braket{T_m}^2=&\tau_m^2+\left(\frac{\tau_{m-1}}{\rho_{m-1}}\right)^2+\left(\frac{\tau_{m+1}}{1-\rho_{m+2}}\right)^2\\
        &-\frac{3}{\left(\frac{\rho_{m-1}}{\tau_{m-1}}+\frac{1-\rho_{m+2}}{\tau_{m+1}}\right)^2}.
    \end{split}
    \label{tau_var}
\end{equation}

We consider the number of events $Q_t$ that particles pass site $m$ until time $t$ to obtain the maximal current and the diffusion coefficient.
For the LD and HD regimes, the density profile is homogeneous on a macroscopic scale.
However, local densities around the target site are fluctuating, i.e., dense or dilute, which affects the passage time.
Therefore, the passage times are not IID random variables for the LD and HD regimes.
For the MC regime, macroscopic density segregation exists.
When the target locates site $m$, particles are constantly dense on the left of the target and dilute on the right.
This segregation does not vary with time.
Therefore, the passage times are considered to be IID random variables for MC regime and
the process of $Q_t$ can be described by a renewal process \cite{Godreche}.
By renewal theory \cite{Godreche}, the mean number of renewals becomes $\braket{Q_t}\sim t/\braket{T_m}$ for $t\rightarrow\infty$.
The current is represented through the mean number of the passing events until time $t$: $J=\lim_{t\rightarrow\infty}\braket{Q_t}/t$.
Thus, we have
\begin{equation}
    J_{\mathrm{max}}\sim\frac{1}{\braket{T_m}}
    \label{J_max}
\end{equation}
for $L\rightarrow\infty$.
The current depends on the disorder realization.
Figure~\ref{fig:QTM_J} shows a good agreement between numerical simulations and the theory.

Using the number of the passing events, we can derive the mean displacement and the variance of the displacement of a tagged particle.
While the tagged particle starting from site $m+1$ returns to site $m+1$, all particles pass between site $m$ and site $m+1$.
Therefore, in the large-$t$ limit, the displacement, $\delta x_t$, is represented by
\begin{equation}
    \delta x_t\sim\frac{LQ_t}{N}=\frac{Q_t}{\rho}.
\end{equation}
By renewal theory \cite{Godreche}, the mean displacement and the variance of the displacement are represented by
\begin{align}
    \langle \delta x_t\rangle&\sim\frac{\langle Q_t\rangle}{\rho}\sim\frac{t}{\rho\langle T_m\rangle},\\
    \begin{split}
        \langle \delta x_t^2\rangle-\langle \delta x_t\rangle^2&\sim\frac{1}{\rho^2}(\langle Q_t^2\rangle-\langle Q_t\rangle^2)\\
        &\sim\frac{1}{\rho^2}\frac{\langle T_m^2\rangle-\langle T_m\rangle^2}{\langle T_m\rangle^3}t
        \label{var_dx}
    \end{split}
\end{align}
for $t\rightarrow\infty$.
Therefore, the diffusion coefficient for the MC regimes is given by
\begin{equation}
    D\sim\frac{1}{\rho^2}\frac{\langle T_m^2\rangle-\langle T_m\rangle^2}{\langle T_m\rangle^3}
    \label{D_MC}
\end{equation}
for $L\rightarrow\infty$.
Figure~\ref{fig:QTM_D} shows a good agreement between numerical simulations and the theory.

We consider that there is a maximum energy trap in the energy landscape.
However, there could be two sites with the same energy, and this energy is the maximum.
We consider this disorder realization.
In Ref.~\cite{Nossan_2013}, the TASEP with two slow sites was studied, and the maximal current depends on the distance between the two slow sites. 
In this model, all sites except the two slow sites have the same rate.
Therefore, the maximal current in our model could depend on the distance between the two sites with the same energy.
However, this disorder realization is a very rare event, so it does not affect our discussions.

\begin{figure}[b]
    \centering
    \subfigure[\hspace{3.5cm}]{\includegraphics[width = 4.5cm]{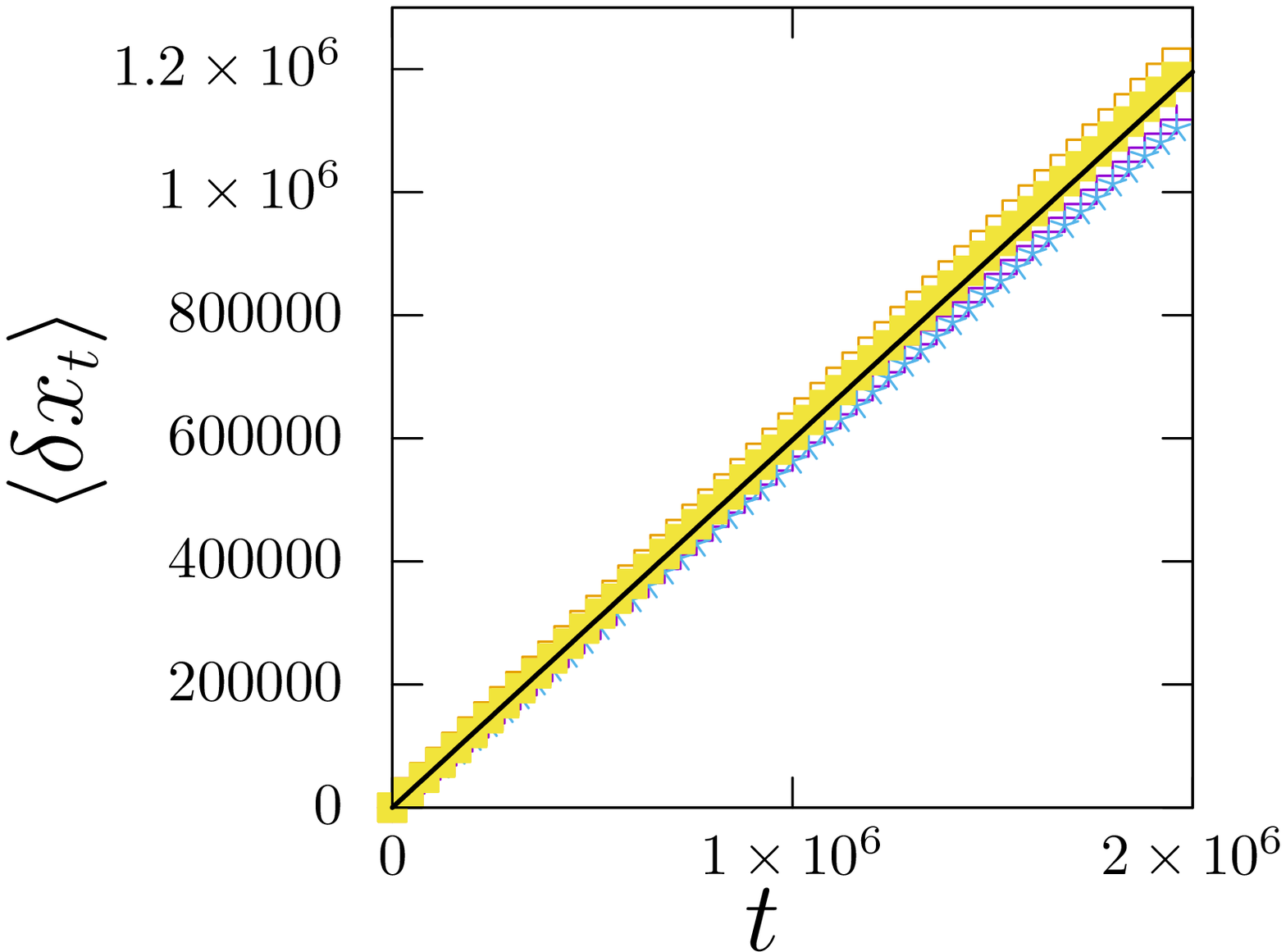}\label{fig:dx_LD}}
    \subfigure[\hspace{3.4cm}]{\includegraphics[width = 4.cm]{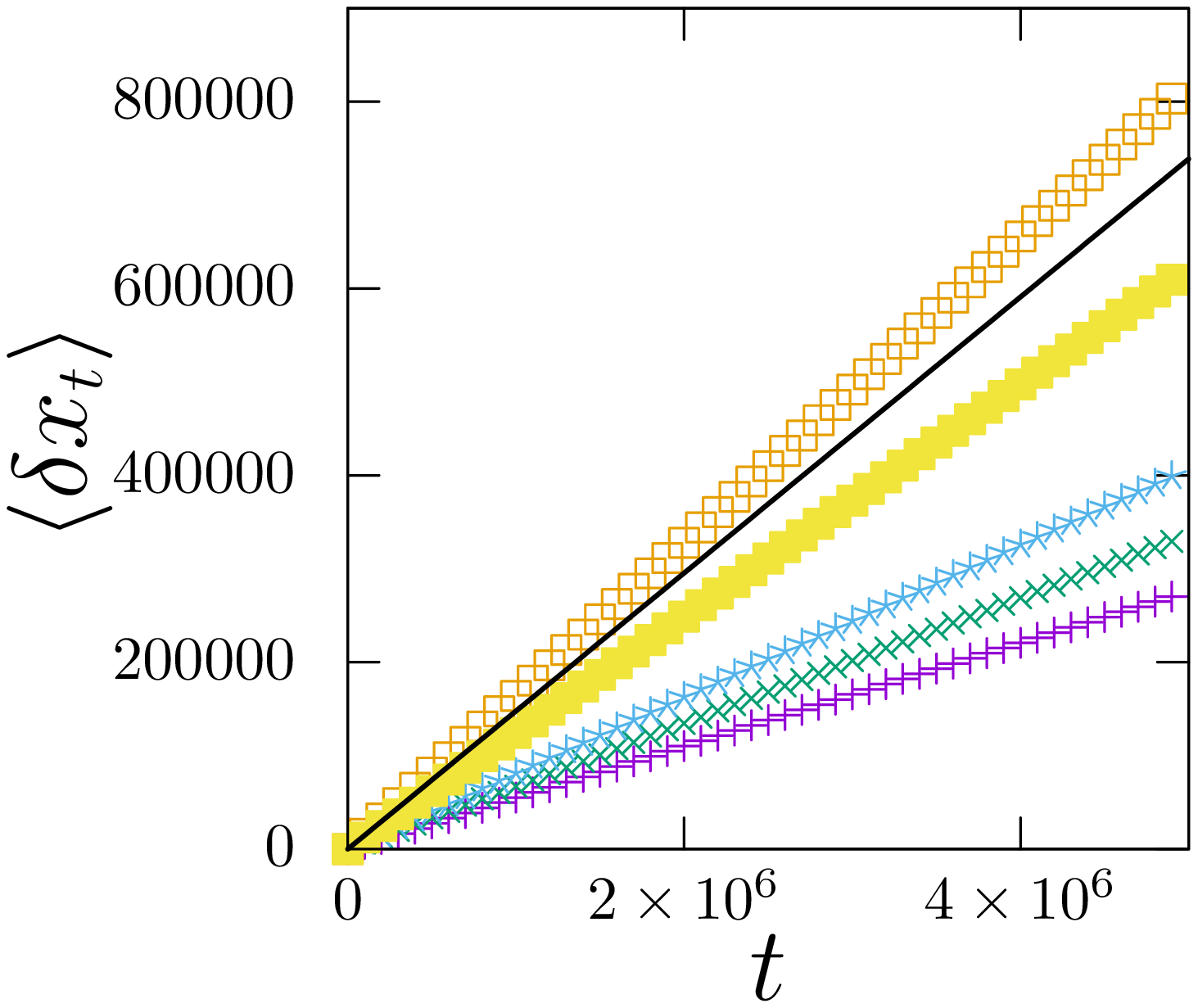}\label{fig:dx_MC}}
    \caption{
    Mean displacement for different density $\rho$: (a) $\rho=0.004$ and (b) $\rho=0.5$ ($\alpha=2.5$ and $L=500$).
    Symbols are the results of mean displacements for five disorder realizations.
    The disorder realizations are the same in both (a) and (b).
    Solid lines represent the disorder averages of the mean displacement, i.e., $\braket{\delta x_t}\sim\braket{J}_{\mathrm{dis}}t/\rho$,
    where $\braket{J}_{\mathrm{dis}}$ was calculated by $\rho(1-\rho)/\braket{\mu}_{\mathrm{dis}}$ for (a), 
    which is calculated by Eq.~(\ref{eq1}), and by Eq.~(\ref{J_dis}) for (b).}
    \label{fig:dx}
\end{figure}
\section{Sample-to-sample fluctuations of current and diffusivity}
\label{SA}
\subsection{Current}
Here, we consider sample-to-sample fluctuations of the current.
To quantify the self-averaging (SA) property of the current, we consider the SA parameter defined as~\cite{AkimotoBarkaiSaito}
\begin{equation}
    \mathrm{SA}(L;J)\equiv \frac{\braket{J(L)^2}_{\mathrm{dis}}-\braket{J(L)}_{\mathrm{dis}}^2}{\braket{J(L)}_{\mathrm{dis}}^2},
    \label{J_SA1}
\end{equation}
where $J(L)$ is the current.
If the SA parameter becomes $0$, there is no sample-to-sample fluctuation, which means SA.

\subsubsection{LD and HD regimes}
\begin{figure}[b]
    \centering
    \subfigure[\hspace{3.8cm}]{\includegraphics[width = 4.35cm]{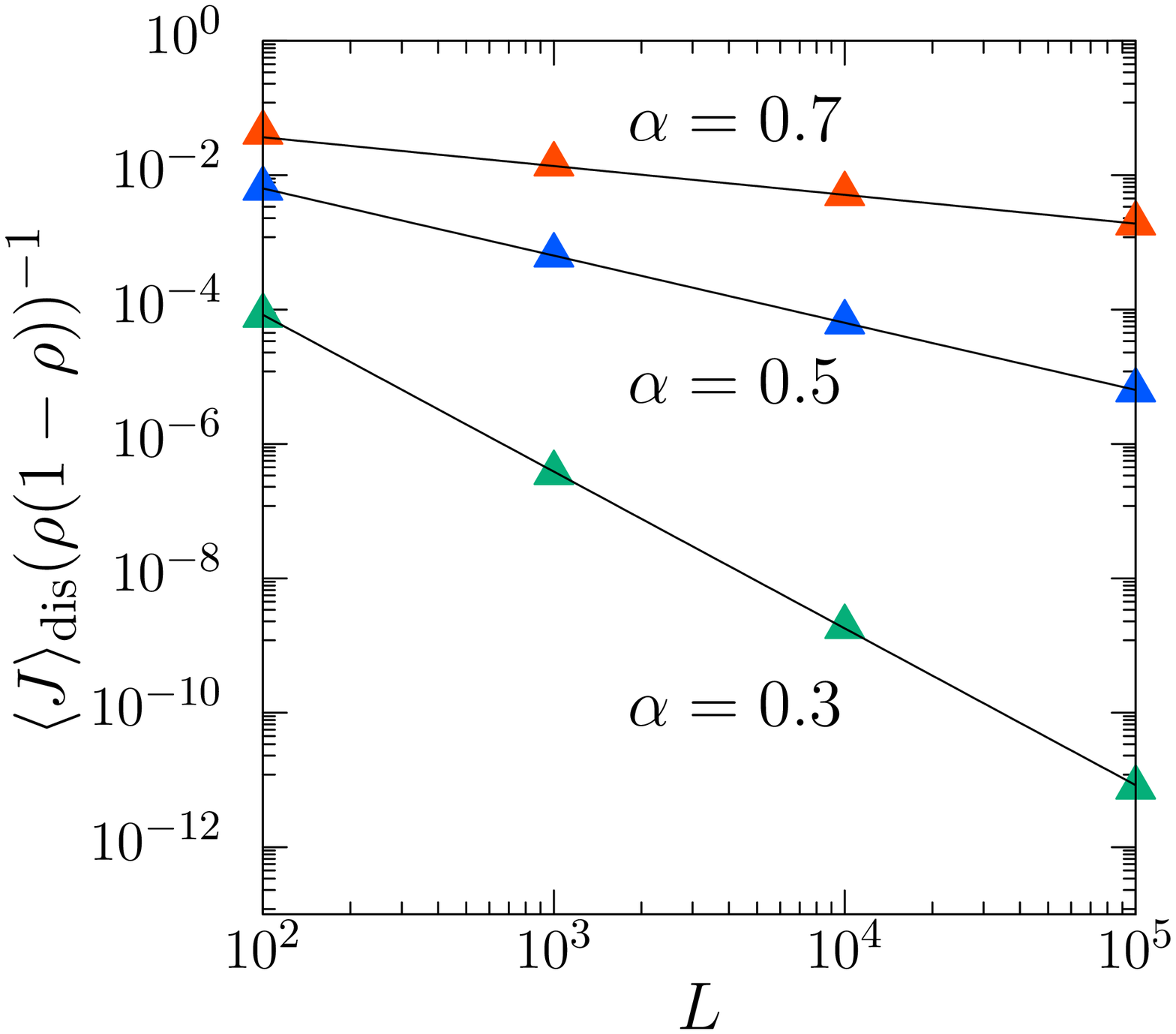}\label{fig:J_LD_av}}
    \subfigure[\hspace{3.5cm}]{\includegraphics[width = 4.2cm]{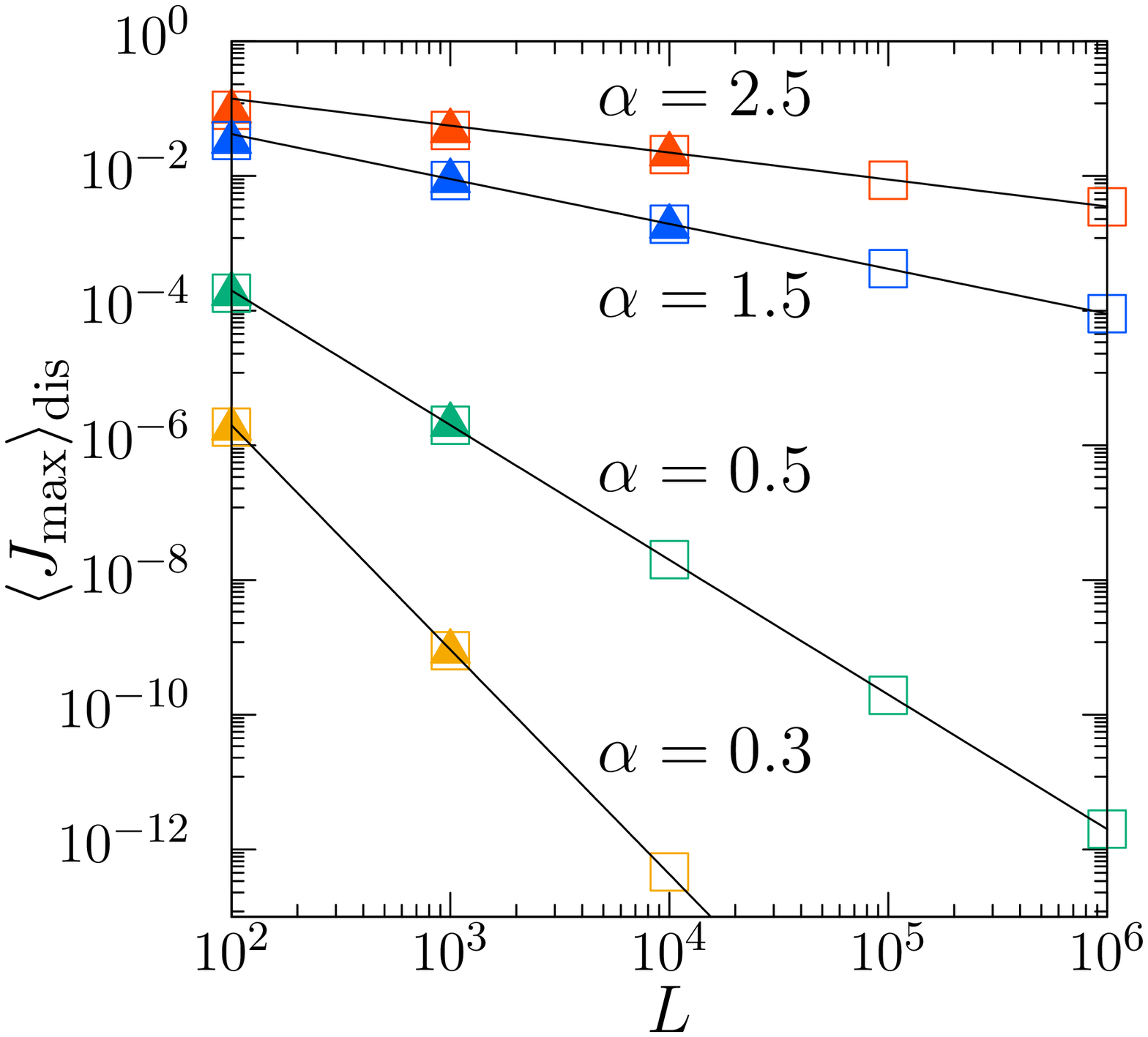}\label{fig:J_av}}
    \caption{Disorder average of the current as a function of $L$ for several $\alpha$: (a) LD and HD regimes and (b) MC regimes.
    Solid lines show the asymptotic results, i.e., Eqs.~(\ref{J_av_LD}) and (\ref{J_dis}).
    Squares are the results of numerical simulations, where we calculated the maximal currents (Eq.~(\ref{J_max}))
    for different disorder realizations by Monte Carlo simulations.
    We used $10^4$ disorder realizations.
    Triangles are the results of the numerical simulation of dynamics of the DTASEP ($N=1$ for (a) and $\rho=0.5$ for (b)).
    We used $10^3$ for $L=10^4$ in the MC regime and $10^4$ disorder realizations for others.}
    \label{fig:QTM_J_av}
\end{figure}

Using Eq.~(\ref{J_LD}), the SA parameter becomes
\begin{equation}
    \mathrm{SA}(L;J)=\frac{\braket{1/\mu^2}_{\mathrm{dis}}-\braket{1/\mu}_{\mathrm{dis}}^2}{\braket{1/\mu}_{\mathrm{dis}}^2},
    \label{J_SA2}
\end{equation}
which is the same as the SA parameter for the diffusion coefficient in the QTM \cite{AkimotoBarkaiSaito}.
When the mean waiting time $\braket{\tau}\equiv \int_0^\infty\tau\psi_\alpha(\tau)d\tau$ is finite ($\alpha>1$), 
we have $\mu\rightarrow\braket{\tau}$ ($L\rightarrow\infty$) by the law of large numbers.
Therefore, in the large-$L$ limit, the current does not depend on the disorder realization (Fig.~\ref{fig:dx}\subref{fig:dx_LD}).
Hence, the current is SA for ${\alpha>1}$.
However, because the scaling of $\rho^*$ follows Eq.~(\ref{rho*}), 
the disorder average of the current in the LD and HD regimes becomes $0$ for $L\rightarrow\infty$.
When the mean waiting time diverges ($\alpha\leq 1$), the law of the large numbers breaks down.
However, the generalized central limit theorem is still valid.
The PDF of the normalized sum of the waiting times follows the one-sided L\'evy distribution \cite{Feller1971},
\begin{equation}
    \frac{\sum_{i=1}^L\tau_i}{L^{1/\alpha}}\Rightarrow X_\alpha\ (L\rightarrow\infty),
\end{equation}
where $X_\alpha$ is a random variable following the one-sided L\'evy distribution of index $\alpha$.
The PDF of $X_\alpha$ denoted by $l_\alpha(x)$ with $x>0$ is given by \cite{Feller1971}
\begin{equation}
    l_\alpha(x)=-\frac{1}{\pi x}\sum_{k=1}^\infty\frac{\Gamma(k\alpha+1)}{k!}(-cx^{-\alpha})^k\sin{(k\pi\alpha)},
\end{equation}
where $c=\Gamma(1-\alpha)\tau_c^\alpha$ is the scale parameter.
The first and the second moment of $X_\alpha^{-1}$ are given by \cite{AkimotoBarkaiSaito}
\begin{equation}
    \langle X_\alpha^{-1}\rangle=\frac{\Gamma(1/\alpha)}{\alpha c^{1/\alpha}},\ 
    \langle X_\alpha^{-2}\rangle=\frac{\Gamma(2/\alpha)}{\alpha c^{2/\alpha}}.
\end{equation}
The current can be represented by
\begin{equation}
    \begin{split}
        J(L)&\sim\rho(1-\rho)\frac{L}{L^{1/\alpha}}\frac{L^{1/\alpha}}{\sum_{k=1}^L\tau_k}\\
        &\sim\rho(1-\rho)L^{1-1/\alpha}X_\alpha^{-1}
    \end{split}
\end{equation}
for $L\rightarrow\infty$.
Thus, the PDF of $J$ is described by the inverse L\'evy distribution.
Using the first moment of the inverse L\'evy distribution \cite{AkimotoBarkaiSaito}, we obtain the exact asymptotic behavior
of the disorder average of the current,
\begin{equation}
    \langle J(L)\rangle_{\mathrm{dis}}\sim\frac{\rho(1-\rho)\Gamma(\alpha^{-1})}{\alpha\tau_c\Gamma(1-\alpha)^{1/\alpha}}L^{1-1/\alpha}.\label{J_av_LD}
\end{equation}
Hence, the current becomes $0$ (see Fig.~\ref{fig:QTM_J_av}\subref{fig:J_LD_av}).
We note that since the scaling of $\rho^*$ follows Eq.~(\ref{rho^*_L}), we do not simulate at the same density.

Using the first and the second moments of $1/\mu$, we have the SA parameter
\begin{equation}
    \lim_{L\rightarrow\infty}\mathrm{SA}(L;J)=
    \left\{\
    \begin{aligned}
        &0 && (\alpha>1)\\
        &\frac{\alpha\Gamma(2/\alpha)}{\Gamma{(1/\alpha)}^2}-1 && (\alpha\leq 1).
    \end{aligned}
    \right.
    \label{J_SA3}
\end{equation}
For $\alpha\leq1$, the SA parameter is a nonzero constant, and thus $J$ becomes non-SA.
Therefore, there is a transition of SA property in the LD and HD regimes.

\subsubsection{MC regime}
When the system size is increased, we find a deeper and deeper energy trap, that is, $\tau_m$ gets longer and longer.
Hence, Eq.~(\ref{tau_ave}) can be approximated as $\langle T_m\rangle\sim\tau_m$, i.e., we can approximate the maximal current:
\begin{equation}
    J_{\max}\sim\frac{1}{\tau_m}.
    \label{J_max2}
\end{equation}
Therefore, the maximal current depend on the disorder realization (Fig.~\ref{fig:dx}\subref{fig:dx_MC}).
Since the PDF of the waiting times follow a power-law distribution Eq.~(\ref{eq1}), 
the PDF of the normalized $\tau_m$ follows the Fr\'echet distribution \cite{HaanFerreira}:
\begin{equation}
    \frac{\tau_m}{\tau_cL^{1/\alpha}}\Rightarrow Y_\alpha\ (L\rightarrow\infty),
    \label{J_dist}
\end{equation}
where $Y_\alpha$ is a random variable following the Fr\'echet distribution of index $\alpha$.
As derived in Appendix \ref{b},
the PDF of $Y_\alpha$, denoted $f_\alpha(y)$ with $y>0$, can be expressed as
\begin{equation}
    f_\alpha(y)=\alpha y^{-\alpha-1}\exp{(-y^{-\alpha})}.
    \label{J_dist2}
\end{equation}
Using Eq.~(\ref{J_dist}), the maximal current can be represented by
\begin{equation}
    J_{\max}(L)\sim\frac{1}{\tau_cL^{1/\alpha}}\frac{\tau_cL^{1/\alpha}}{\tau_m}\sim\frac{1}{\tau_cL^{1/\alpha}}Y_\alpha^{-1}
    \label{J_dist3}
\end{equation}
for $L\rightarrow\infty$. Thus, the PDF of $J_{\max}$ is described by the inverse Fr\'echet distribution.

The PDF of $Y_\alpha^{-1}$ can be explicitly represented by the Fr\'echet distribution:
\begin{equation}
    \Pr(Y_\alpha^{-1}\leq z)=\Pr(Y_\alpha\geq z^{-1})=\int_{z^{-1}}^{\infty}f_\alpha(y)dy.
\end{equation}
The distribution is the Weibull distribution.
We obtain the PDF of $Y_\alpha^{-1}$, denoted by $w_\alpha(z)$:
\begin{equation}
    w_\alpha(z)=\alpha z^{\alpha-1}\exp{(-z^\alpha)}.
\end{equation}
The first and second moments of the Weibull distribution are given by
\begin{equation}
    \braket{Y_\alpha^{-1}}=\Gamma\left(1+\frac{1}{\alpha}\right),\ \ \braket{Y_\alpha^{-2}}=\Gamma\left(1+\frac{2}{\alpha}\right).
    \label{J_ave}
\end{equation}
From Eq.~(\ref{J_ave}), we obtain the exact asymptotic behavior of the disorder average of the maximal current,
\begin{equation}
    \braket{J_{\max}(L)}_{\mathrm{dis}}\sim\frac{1}{\tau_cL^{1/\alpha}}\Gamma\left(1+\frac{1}{\alpha}\right).
    \label{J_dis}
\end{equation}
Therefore, the maximal current decreases with the system size $L$ (see Fig.~\ref{fig:QTM_J_av}\subref{fig:J_av}).

\begin{figure*}[t]
    \subfigure[\hspace{8cm}]{\includegraphics[width = 8.6cm]{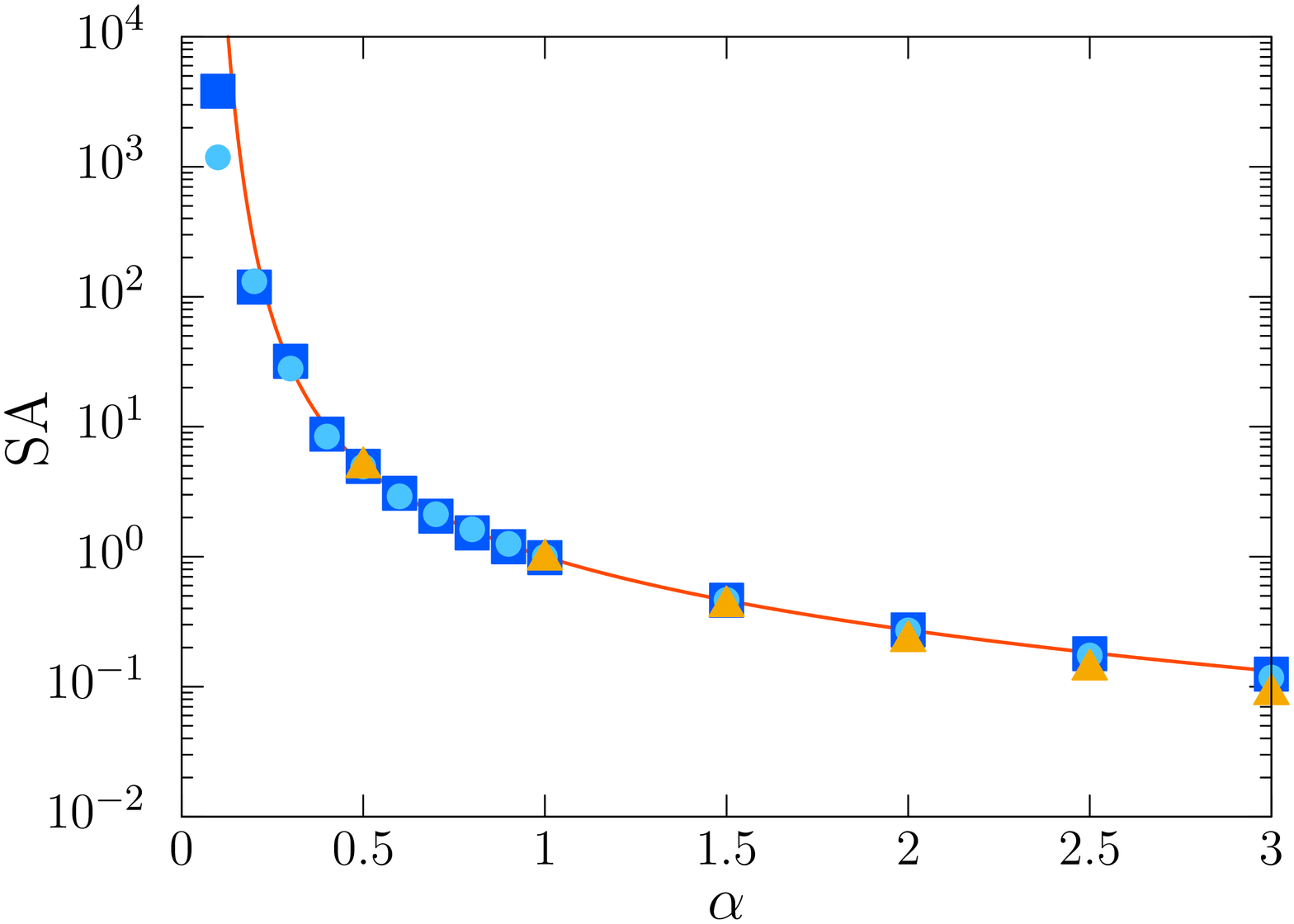}\label{fig:J_D_SA}}\\
    \subfigure[\hspace{8cm}]{\includegraphics[width = 8.6cm]{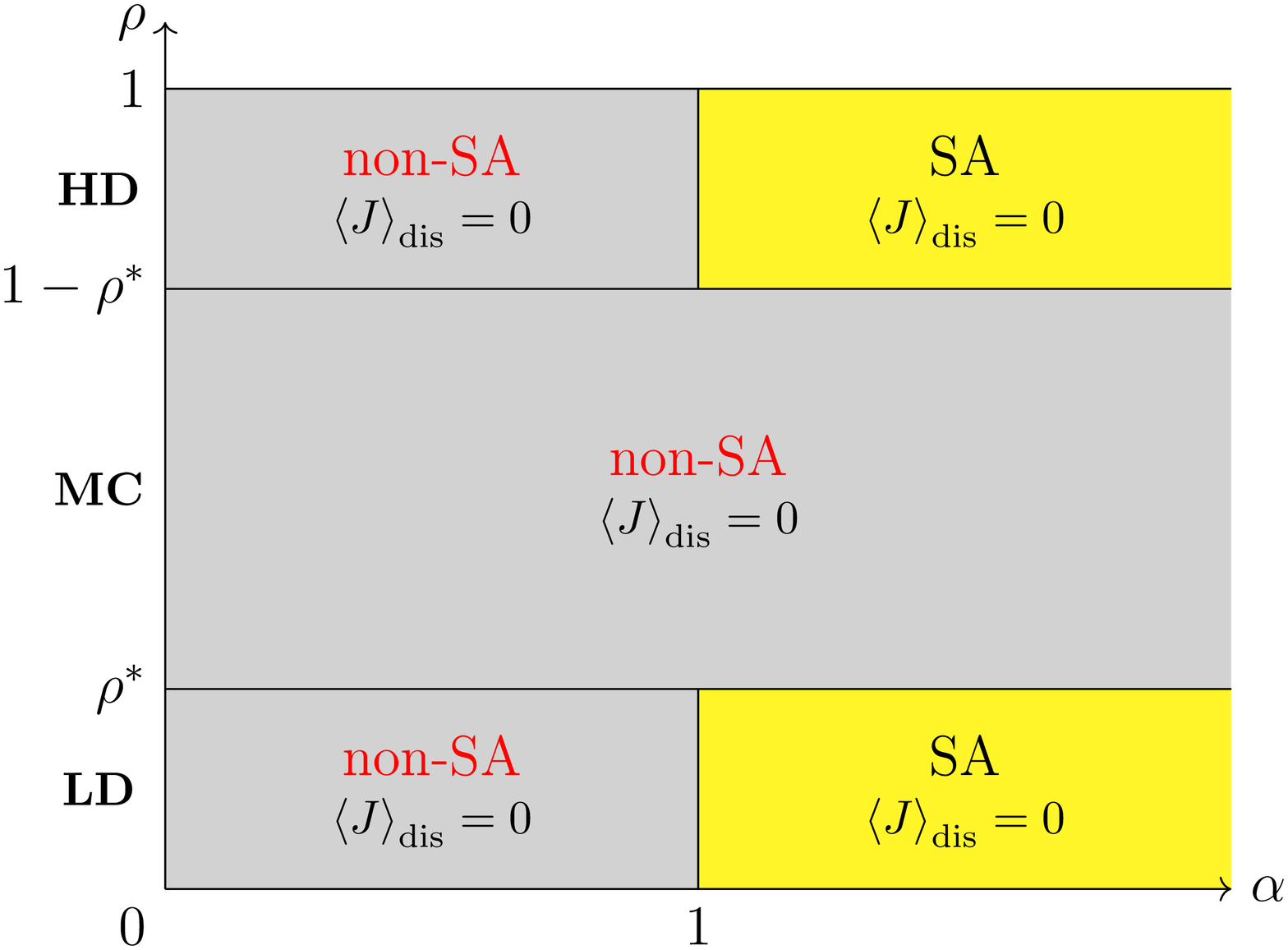}\label{fig:J_SA}}
    \subfigure[\hspace{8cm}]{\includegraphics[width = 8.6cm]{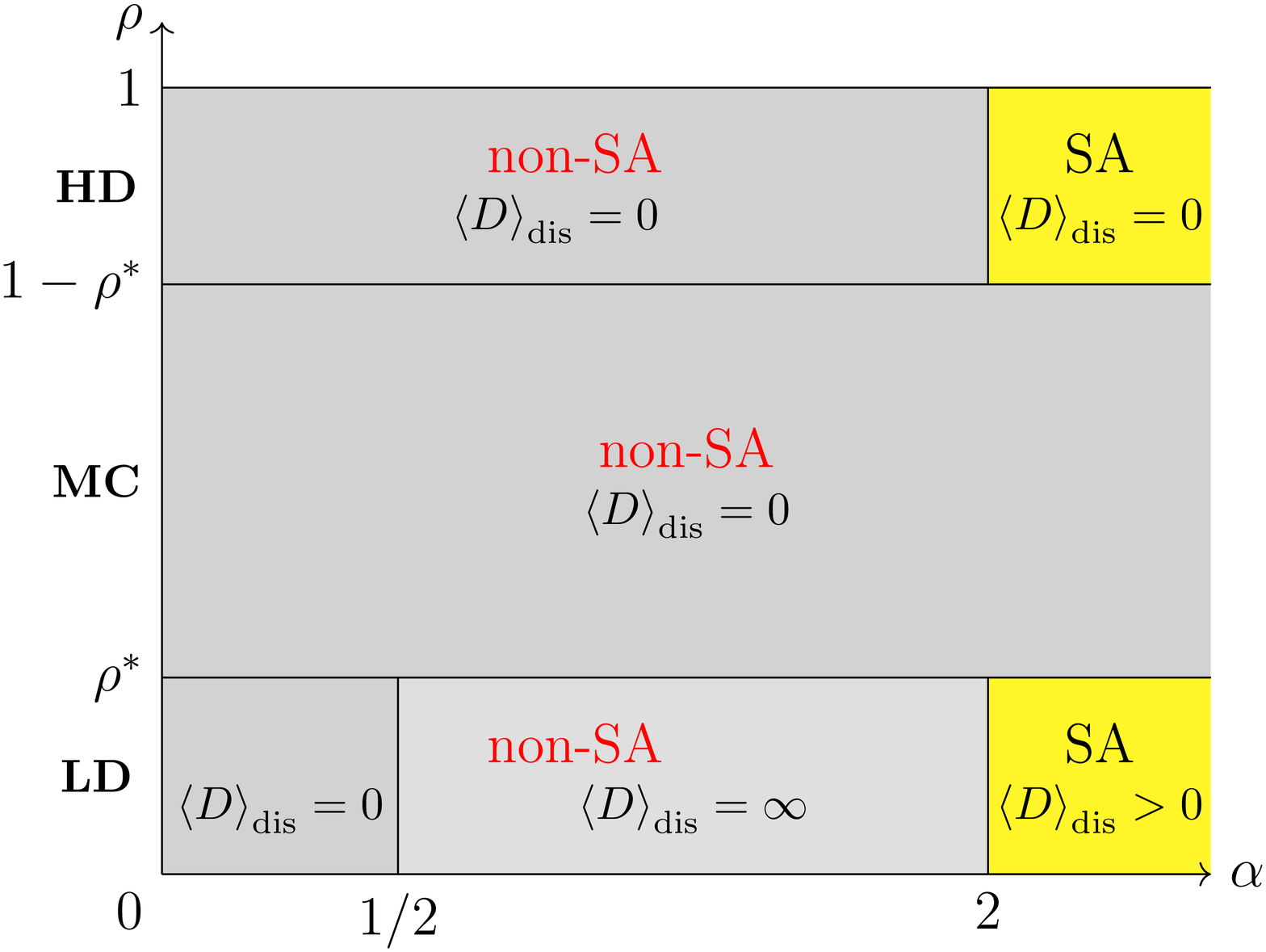}\label{fig:D_SA}}
    \caption{(a) Self-averaging parameter as a function of $\alpha$.
        The squares and circles are the results of numerical simulations, where we calculated the maximal currents (Eq.~(\ref{J_max}))
        and the diffusion coefficient (Eq.~(\ref{D_MC})) for different disorder realizations by Monte Carlo simulations ($L=10^5$), respectively.
        The triangles show the self-averaging parameter of the maximal current
        obtained by the numerical simulation of the dynamics of the DTASEP ($L=1000$ and $N=500$).
        We used $10^4$ disorder realizations.
        The solid line represents Eq.~(\ref{J_max_SA2}).
        (b) Phase diagram based on current in the LD, MC, and HD regimes.
        (c) Phase diagram based on diffusivity in the LD, MC, and HD regimes.
        }
    \label{fig:QTM_SA}
\end{figure*}
Let us consider the SA property for the maximal current.
The SA parameter is defined as
\begin{equation}
    \mathrm{SA}(L;J_{\mathrm{max}})\equiv \frac{\braket{J_{\max}(L)^2}_{\mathrm{dis}}-\braket{J_{\max}(L)}_{\mathrm{dis}}^2}{\braket{J_{\max}(L)}_{\mathrm{dis}}^2}.
    \label{J_max_SA1}
\end{equation}
Using Eq.~(\ref{J_dist3}), we have
\begin{equation}
    \begin{split}
        \lim_{L\rightarrow\infty}\mathrm{SA}(L;J_{\mathrm{max}})
        &=\frac{\braket{Y_\alpha^{-2}}-\braket{Y_\alpha^{-1}}^2}{\braket{Y_\alpha^{-1}}^2}\\
        &=\frac{\Gamma\left(1+2/\alpha\right)}{\Gamma\left(1+1/\alpha\right)^2}-1.
    \end{split}
    \label{J_max_SA2}
\end{equation}
The SA parameter becomes a nonzero constant, i.e., the maximal current becomes non-SA (see Fig.~\ref{fig:QTM_SA}\subref{fig:J_D_SA}).
This result differs from LD and HD, and there is no transition from SA to non-SA behavior for all $\alpha$ (see Fig.~\ref{fig:QTM_SA}\subref{fig:J_SA}).
As shown in Fig.~\ref{fig:dx}, the currents for different disorder realizations exhibit non-SA in the MC regime, 
whereas they are SA in the LD regime even when the disorder realizations are the same in both regimes.
Therefore, this is clear evidence of the many-body effect in the DTASEP.

\begin{figure}[t]
    \centering
    \subfigure[\hspace{3.5cm}]{\includegraphics[width = 4.2cm]{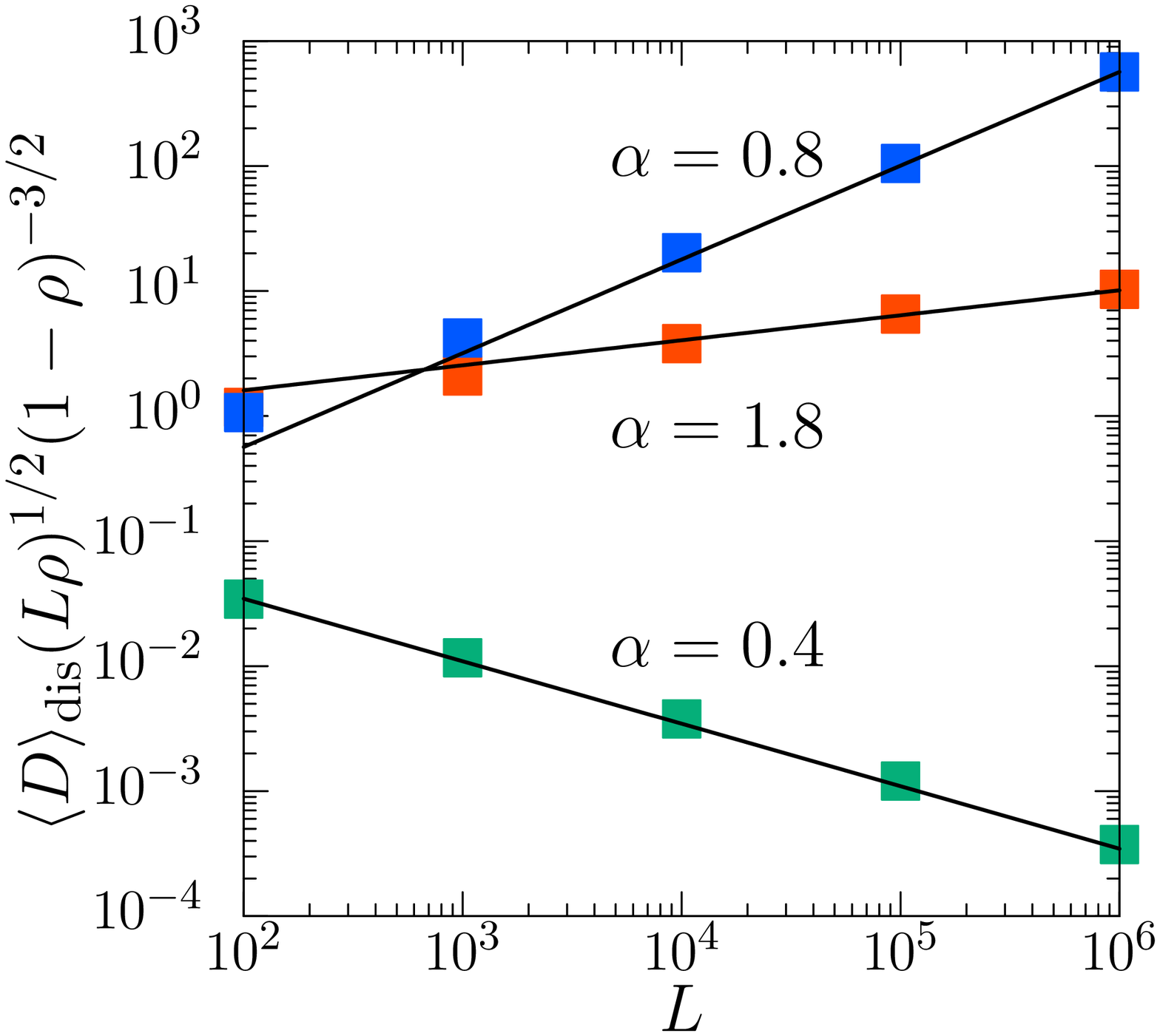}\label{fig:D_LD_av}}
    \subfigure[\hspace{3.5cm}]{\includegraphics[width = 4.2cm]{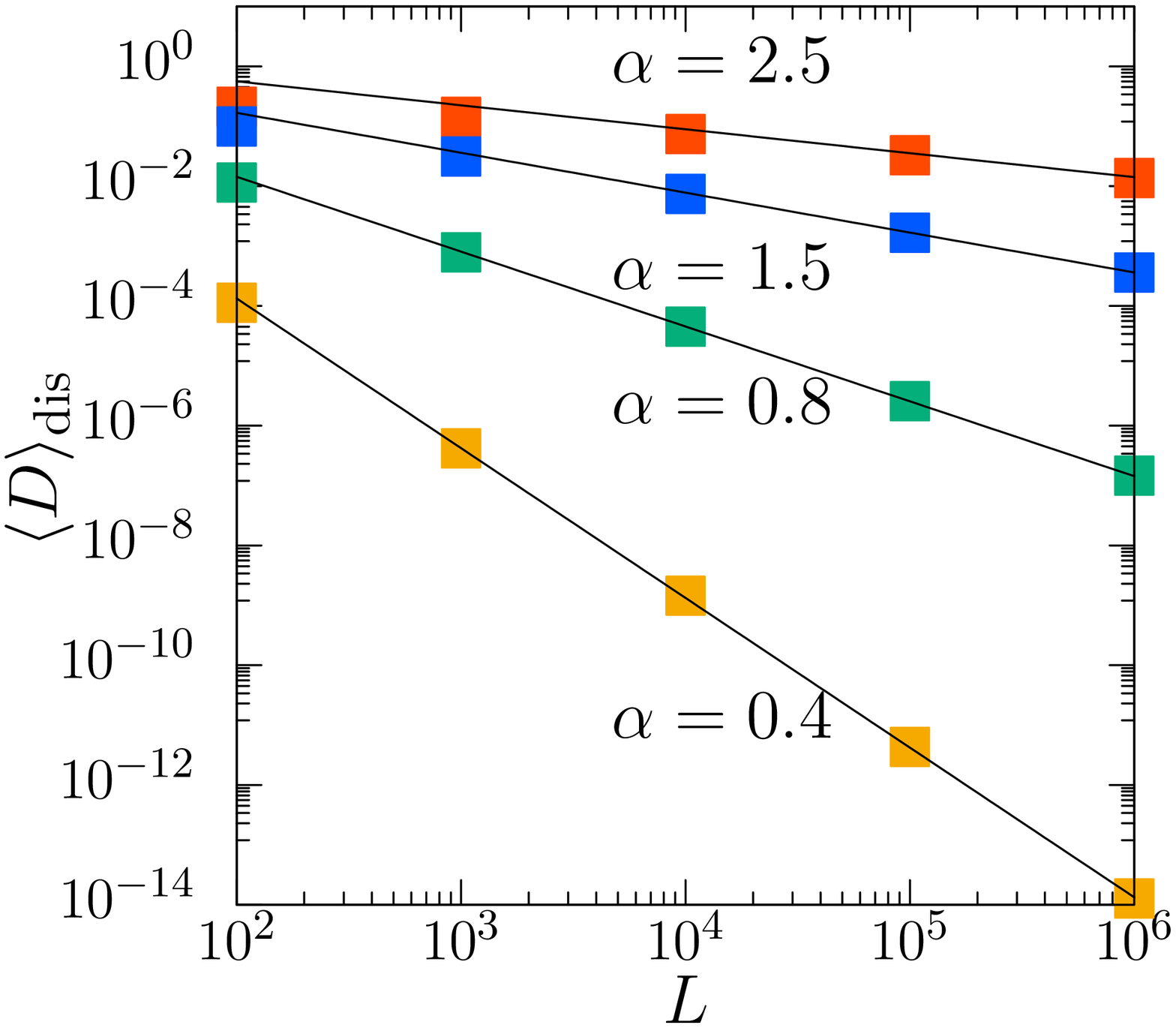}\label{fig:D_av}}
    \caption{Disorder average of the diffusion coefficient as a function of $L$ for several $\alpha$: (a) LD and HD regimes and (b) MC regimes.
    Squares are the results of numerical simulations, where we calculated the diffusion coefficient (Eqs.~(\ref{D_LD}) and (\ref{D_MC}))
    for different disorder realizations by Monte Carlo simulations ($N=1$ for (a) and $\rho=0.5$ for (b)).
    We used $10^4$ disorder realizations.
    Solid lines show the asymptotic results, i.e., Eqs.~(\ref{D_av}) and (\ref{D_dis}).}
    \label{fig:QTM_D_av}
\end{figure}

\subsection{Diffusivity}
Here, we consider sample-to-sample fluctuations of the diffusion coefficient.
In the homogeneous TASEP, the diffusion coefficient becomes $0$ for $L\rightarrow\infty$ (Eq.~(\ref{TASEP2}))
because of the many-body effect.
$D=0$ in the homogeneous TASEP on a finite system implies the subdiffusion in that on an infinite system \cite{Beijeren:1991aa}.
\subsubsection{LD and HD regimes}
For the LD regime, $\rho=N/L$ and $1-\rho\sim 1$ for $L\rightarrow\infty$ and $N\ll L$.
We define the number of holes as ${M\equiv L-N}$, i.e., $1-\rho=M/L$.
Therefore, for the HD regime,  $\rho=(L-M)/L\sim 1$ for $L\rightarrow\infty$ and $M\ll L$.
Using Eq.~(\ref{D_LD}), the disorder average of the diffusion coefficient is given by
\begin{equation}
    \Braket{D(L)}_{\mathrm{dis}}\sim
    \left\{\
    \begin{aligned}
        &N^{-1/2}\Braket{\frac{\sigma^2}{\mu^3}}_{\mathrm{dis}} && (LD\ regime)\\
        &M^{3/2}L^{-2}\Braket{\frac{\sigma^2}{\mu^3}}_{\mathrm{dis}} && (HD\ regime)
    \end{aligned}
    \right.
\end{equation}
for $L\rightarrow\infty$.
When the second moment of the waiting time $\braket{\tau^2}\equiv \int_0^\infty\tau^2\phi_\alpha(\tau)d\tau$ is finite ($\alpha>2$),
we have $\sigma^2\rightarrow\braket{\tau^2}$ ($L\rightarrow\infty$) by the law of large numbers.
It follows that the disorder average of $D(L)$ is finite and given by
\begin{equation}
    \Braket{D(L)}_{\mathrm{dis}}\sim
    \left\{\
    \begin{aligned}
        &N^{-1/2}\frac{\braket{\tau^2}}{\braket{\tau}^3} && (LD\ regime)\\
        &M^{3/2}L^{-2}\frac{\braket{\tau^2}}{\braket{\tau}^3} && (HD\ regime)
    \end{aligned}
    \right.
\end{equation}
for $L\rightarrow\infty$ and $\alpha>2$.
Hence, the diffusion coefficient become non-zero constant for the LD regime,
whereas it becomes $0$ for the HD regime.

For $\alpha<2$, the second moment of the waiting time diverges.
The disorder average of $\sigma^2/\mu^3$, which was derived in Ref.~\cite{AkimotoSaito2020},
is obtained as
\begin{equation}
    \Braket{\frac{\sigma^2}{\mu^3}}_{\mathrm{dis}}\propto
    \left\{\
    \begin{aligned}
        &L^{2-\alpha} && (1<\alpha<2)\\
        &L^{2-1/\alpha} && (\alpha< 1).
    \end{aligned}
    \right.
\end{equation}
Therefore, the disorder average of the diffusion coefficient is given by
\begin{equation}
    \Braket{D(L)}_{\mathrm{dis}}\propto 
    \left\{\
    \begin{aligned}
        &L^{2-\alpha} && (1<\alpha<2)\\
        &L^{2-1/\alpha} && (\alpha<1)
    \end{aligned}
    \right.\label{D_av}
\end{equation}
for the LD regime and
\begin{equation}
    \Braket{D(L)}_{\mathrm{dis}}\propto 
    \left\{\
    \begin{aligned}
        &L^{-\alpha} && (1<\alpha<2)\\
        &L^{-1/\alpha} && (\alpha<1)
    \end{aligned}
    \right.
\end{equation}
for the HD regime, respectively.
Hence, the diffusion coefficient for the LD regime diverges for $1<\alpha<2$ and $1/2<\alpha<1$,
whereas it becomes $0$ for $\alpha<1/2$ (see Fig.~\ref{fig:QTM_D_av}\subref{fig:D_LD_av}).
The diffusion coefficient for the HD regime becomes $0$ for all $\alpha$.
The zero diffusion coefficient is a signature of many-body effect.

Let us consider the SA property for the diffusion coefficient in LD and HD regimes.
The SA parameter is defined as
\begin{equation}
    \mathrm{SA}(L;D)\equiv \frac{\braket{D(L)^2}_{\mathrm{dis}}-\braket{D(L)}_{\mathrm{dis}}^2}{\braket{D(L)}_\mathrm{dis}^2}.
\end{equation}
The SA parameter goes to $0$ in the large-$L$ limit when the diffusion coefficient is SA.

For $\alpha>2$, the second moment of waiting times exists; i.e., $\braket{\tau^2}=\int_0^\infty \tau^2\psi_\alpha(\tau)d\tau$.
Thus, $\sigma^2/\mu^3$ converges to $\braket{\tau^2}/\braket{\tau}^3$ for $L\rightarrow\infty$.
Therefore, $\braket{D(L)^2}_{\mathrm{dis}}-\braket{D(L)}_{\mathrm{dis}}^2$ converges to $0$ for $L\rightarrow\infty$,
so that the diffusion coefficient is SA for $\alpha>2$.

For $1<\alpha<2$, the second moment of $\sigma^2/\mu^3$ was calculated in Ref.~\cite{AkimotoSaito2020}.
The SA parameter diverges as
\begin{equation}
    \mathrm{SA}(L;D)\propto\frac{\braket{D(L)^2}_{\mathrm{dis}}}{\braket{D(L)}_{\mathrm{dis}}^2}\propto L^{\alpha-1}
\end{equation}
for $L\rightarrow\infty$. Therefore, the diffusion coefficient is non-SA for $1<\alpha<2$.

For $\alpha<1$, both the first and the second moments of the waiting times diverge.
$\sigma^2/\mu^3$ can be represented as 
\begin{equation}
    \frac{\sigma^2}{\mu^3}=L^{2-1/\alpha}C(L),
\end{equation}
where $C(L)=L^{1/\alpha}\sum_{i=1}^L\tau_i^2/(\sum_{i=1}^L\tau_i)^3$ is a random variable depending on the disorder realization.
Hence, the SA parameter becomes
\begin{equation}
    \mathrm{SA}(L;D)=\frac{\braket{D(L)^2}_{\mathrm{dis}}}{\braket{D(L)}_{\mathrm{dis}}^2}-1
    =\frac{\braket{C(L)^2}_{\mathrm{dis}}}{\braket{C(L)}_{\mathrm{dis}}^2}-1.
\end{equation}
Because ${\sum_{i=1}^L\tau_i^2<(\sum_{i=1}^L\tau)^3}$, $1/(\sum_{i=1}^L\tau_i)^3<C(L)<1$, i.e., 
$0<\braket{C(L)}_{\mathrm{dis}}<1$, and $0<\braket{C(L)^2}_{\mathrm{dis}}<1$,
the SA  parameter is a finite value, i.e., the diffusion coefficient is non-SA for $\alpha<1$.
These results are the same as those for the QTM.

\subsubsection{MC regime}
When the system size is increased, we find a deeper and deeper energy trap, that is, $\tau_m$ gets longer and longer.
Hence, Eq.~(\ref{tau_var}) can be approximated as $\langle T_m^2\rangle-\langle T_m\rangle^2\sim\tau_m^2$, i.e.,
we can approximate the diffusion coefficient:
\begin{equation}
    D\sim\frac{\rho^{-2}}{\tau_m}.
\end{equation}
By Eq.~(\ref{J_dist}), the diffusion coefficient can be represented by
\begin{equation}
    D(L)\sim\frac{\rho^{-2}}{\tau_cL^{1/\alpha}}\frac{\tau_cL^{1/\alpha}}{\tau_m}\sim\frac{\rho^{-2}}{\tau_cL^{1/\alpha}}Y_\alpha^{-1}
    \label{D_dist}
\end{equation}
for $L\rightarrow\infty$.
Therefore, the PDF of the diffusion coefficient is also described by the Weibull distribution. 
Using the first moment of the Weibull distribution, we obtain the exact asymptotic behavior of the disorder average of 
the diffusion coefficient,
\begin{equation}
    \braket{D(L)}_{\mathrm{dis}}\sim\frac{\rho^{-2}}{\tau_cL^{1/\alpha}}\Gamma(1+1/\alpha).
    \label{D_dis}
\end{equation}
Therefore, the diffusion coefficient also decreases with the system size $L$ (see Fig.~\ref{fig:QTM_D_av}\subref{fig:D_av}).

Next, we consider the SA parameter of the diffusion coefficient in the MC regime.
Using Eq.~(\ref{D_dist}), we have
\begin{equation}
    \begin{split}
        \lim_{L\rightarrow\infty}\mathrm{SA}(L;D)&=\frac{\braket{Y_\alpha^{-2}}-\braket{Y_\alpha^{-1}}^2}{\braket{Y_\alpha^{-1}}^2}\\
        &=\frac{\Gamma\left(1+2/\alpha\right)}{\Gamma\left(1+1/\alpha\right)^2}-1,
    \end{split}
\end{equation}
which is the same as the SA parameter for the maximal current (see Fig.~\ref{fig:QTM_SA}\subref{fig:J_D_SA}).
The transition point from SA to non-SA, which exists for the LD and HD regimes, disappears, 
and the diffusion coefficient is non-SA for all $\alpha$ (see Fig.~\ref{fig:QTM_SA}\subref{fig:D_SA}).

\section{Conclusion}\label{conclusion}
In this paper, we have studied the TASEP on a quenched random energy landscape.
In the LD and HD regimes, i.e., the dilute limit, the dynamics of the disordered TASEP can be approximately described by the single-particle dynamics.
On the other hand, the dynamics in the MC regime become completely different from that in the dilute limit due to the many-body effect.
In particular, the LD and HD phases coexist in the MC regimes.
By renewal theory, we provided exact results for the current and diffusion coefficient.
In the LD regime, the disorder average of the diffusion coefficient becomes $0$ for $\alpha<1/2$, diverges for $1/2<\alpha<2$,
and is non-zero constant for $\alpha>2$, which is the same as in the single-particle dynamics (Fig.~\ref{fig:QTM_SA}\subref{fig:D_SA}).
On the other hand, in the HD and MC regimes, it becomes $0$ in the large-$L$ limit for all $\alpha$ 
(Fig.~\ref{fig:QTM_SA}\subref{fig:D_SA}) due to the many-body effect.
Moreover, we introduced the SA parameter to quantify the SA property.
We obtained a self-averaging and non-self-averaging transition for the current and the diffusion coefficient in the LD and HD regimes, 
which is the same as in the single-particle dynamics.
However, in the MC regime, the current and diffusion coefficient are non-SA for all $\alpha$,
which is different from the LD and HD regimes.
Therefore, many-body effects in quenched random energy landscapes decrease the diffusion coefficient and lead to a strong non-self-averaging feature.

\begin{acknowledgments}
    We thank K. Saito for fruitful discussions.
    T.A. was supported by JSPS Grant-in-Aid for Scientific Research (No. C JP21K033920).
\end{acknowledgments}

\appendix
\section{Passage time distribution}
\label{a}
\begin{figure}[b]
    \centering
    \includegraphics[width=8.6cm]{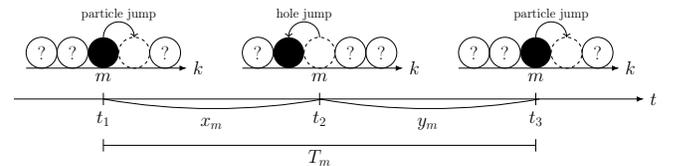}
    \caption{Particle dynamics during the passage time.
    The filled and dashed-line circles denote particles and holes, respectively.
    The question mark is either a particle or a hole.}
    \label{fig:A1}
\end{figure}
In this appendix, we derive the distribution of the passage time $T_m$ site $m$ in the MC regime,
where $m$ is the site with the maximal mean waiting time.
The passage time can be divided into the hole escape time $x_m$ and the particle escape time $y_m$.
At time $t_1$, a particle escapes from site $m$.
At time $t_2$, the subsequent particle arrives at site $m$.
The hole escape time is defined as $x_m=t_2-t_1$ (Fig.~\ref{fig:A1}).
At time $t_3$, the particle escapes from site $m$.
The particle escape time is defined as $y_m=t_3-t_2$ (Fig.~\ref{fig:A1}).
To obtain the hole escape time at site $m$, we consider the hole dynamics.
At site $m$, when the hole jump succeeds by the $i$th attempt, the PDF of the hole escape time $x_m$ follows
the distribution of the sum of $i$ IID variables following the exponential distribution, $\psi_{m-1}(t)=\tau_{m-1}^{-1}\exp{(-t/\tau_{m-1})}$, 
i.e., the Erlang distribution
\begin{equation}
    Er(x_m;i,\tau_{m-1})=\frac{x_m^{i-1}}{(i-1)!\tau_{m-1}^i}\exp{\left(-\frac{x_m}{\tau_{m-1}}\right)},
\end{equation}
and the success probability is given by $\rho_{m-1}(1-\rho_{m-1})^{i-1}$.
Therefore, the PDF $f(x_m)$ of $x_m$ follows the exponential distribution
\begin{equation}
    \begin{split}
        f(x_m)=&\rho_{m-1}\sum_{i=1}^\infty(1-\rho_{m-1})^{i-1}Er(x_m;i,\tau_{m-1})\\
        =&\frac{\rho_{m-1}}{\tau_{m-1}}\exp{\left(-\frac{x_m}{\tau_{m-1}}\right)}\\
        &\times\sum_{i=1}^\infty\frac{1}{(i-1)!}\left(\frac{(1-\rho_{m-1})x_m}{\tau_{m-1}}\right)^{i-1}\\
        =&Ex\left(x_m;\frac{\tau_{m-1}}{\rho_{m-1}}\right),
    \end{split}
    \label{A2}
\end{equation}
where $Ex(x;\tau)\equiv\exp{(-x/\tau)}/\tau$ is the exponential distribution.

Because a particle can not escape from site $m$ until the neighbor site becomes empty, we must consider the effect of site $m+1$.
Using the same way of the derivation of Eq.~(\ref{A2}), the PDF $g(y_{m+1})$ of the particle escape time $y_{m+1}$ at site $m+1$ is given by
\begin{equation}
    g(y_{m+1})=Ex\left(y_{m+1};\frac{\tau_{m+1}}{1-\rho_{m+2}}\right).
    \label{A3}
\end{equation}
Using Eq.~(\ref{A3}), we derive the joint PDF of the hole escape time $x_m$ and the particle escape time $y_m$.
When the sum of the hole escape time $x_m$ and the particle escape time $y_m$ is larger than the particle escape time $y_{m+1}$,
a particle at site $m$ can jump to site $m+1$.
When a particle succeeds to jump to site $m+1$ once, i.e., $x_m+y_m>y_{m+1}$, the weighted joint PDF $h_1(x_m,y_m)$ of $x_m$ and $y_m$
is given by
\begin{equation}
    h_1(x_m,y_m)=f(x_m)Ex(y_m;\tau_m)\int_0^{x_m+y_m}dy_{m+1}\ g(y_{m+1}).
\end{equation}
When a particle jump succeeds on the $i$th attempts ($i>1$), $x_m+y_m'<y_{m+1}<x_m+y_m$,
where $y_m'$ follows the Erlang distribution $Er(y_m';i-1,\tau_m)$ and
$y_m$ is sum of $y_m'$ and the IID random variable $y$ with the exponential distribution $Ex(y;\tau_m)$.
Then, the weighted joint PDf $h_i(x_m,y_m)$ of $x_m$ and $y_m$ is given by
\begin{widetext}
\begin{equation}
    h_i(x_m,y_m)=f(x_m)\int_0^{y_m}dy_m'\ Ex(y_m-y_m';\tau_m)Er(y_m';i-1,\tau_m)\int_{x_m+y_m'}^{x_m+y_m}dy_{m+1}\ g(y_{m+1}).
\end{equation}
Therefore, the joint PDF $h(x_m,y_m)$ of $x_m$ and $y_m$ is given by
\begin{equation}
    \begin{split}
        h(x_m,y_m)&=\sum_{i=1}^\infty h_i(x_m,y_m)\\
        &=f(x_m)Ex(y_m;\tau_m)+\frac{\frac{\tau_{m+1}}{1-\rho_{m+2}}}{\tau_m-\frac{\tau_{m+1}}{1-\rho_{m+2}}}\exp{\left(-\frac{1-\rho_{m+2}}{\tau_{m+1}}x_m\right)}
        f(x_m)[Ex(y_m;\tau_m)-g(y_m)].
    \end{split}
\end{equation}
By the convolutional integration of $h(x_m,y_m)$, we have the PDF $\Phi(T_m)$ of the passage time $T_m$
\begin{equation}
    \begin{split}
        \Phi(T_m)&=\int_0^{T_m}dx\ h(x,T_m-x)\\
        &=\tau_m\frac{\rho_{m-1}}{\tau_{m-1}}(\zeta_1+\zeta_2\zeta_3)Ex(T_m;\tau_m)-\zeta_1f(T_m)-\zeta_2g(T_m)+\zeta_3E\left(T_m;\frac{1}{\frac{\rho_{m-1}}{\tau_{m-1}}+\frac{1-\rho_{m+2}}{\tau_{m+1}}}\right),
    \end{split}
\end{equation}
where
\begin{equation*}
    \zeta_1\equiv\frac{1}{\tau_m\frac{\rho_{m-1}}{\tau_{m-1}}-1},\ \zeta_2\equiv\frac{1}{\tau_m\frac{1-\rho_{m+2}}{\tau_{m+1}}-1},\ 
    \zeta_3\equiv\frac{1}{\tau_m\left(\frac{\rho_{m-1}}{\tau_{m-1}}+\frac{1-\rho_{m+2}}{\tau_{m+1}}\right)-1}.
\end{equation*}

Next, we derive the mean and variance of the passage time.
The Laplace transform of $\Phi(T_m)$ with respect to $s$ is given by
\begin{equation}
    \begin{split}
        \hat{\Phi}(s)&\equiv\int_0^\infty dT_m\ e^{-sT_m}\Phi(T_m)\\
        &=\tau_m\frac{\rho_{m-1}}{\tau_{m-1}}(\zeta_1+\zeta_2\zeta_3)\frac{1}{\tau_ms+1}-\frac{\zeta_1}{\frac{\tau_{m-1}}{\rho_{m-1}}s+1}-\frac{\zeta_2}{\frac{\tau_{m+1}}{1-\rho_{m+2}}s+1}+\frac{\zeta_3}{\frac{s}{\frac{\rho_{m-1}}{\tau_{m-1}}+\frac{1-\rho_{m+2}}{\tau_{m+1}}}+1}.
    \end{split}
\end{equation}
It follows that the mean and variance of the passage time are given by
\begin{align}
    &\braket{T_m}=\tau_m+\frac{\tau_{m-1}}{\rho_{m-1}}+\frac{\frac{\rho_{m-1}}{\tau_{m-1}}}{\frac{\rho_{m-1}}{\tau_{m-1}}+\frac{1-\rho_{m+2}}{\tau_{m+1}}}\frac{\tau_{m+1}}{1-\rho_{m+2}},\\
    &\braket{T_m^2}-\braket{T_m}^2=\tau_m^2+\left(\frac{\tau_{m-1}}{\rho_{m-1}}\right)^2+\left(\frac{\tau_{m+1}}{1-\rho_{m+2}}\right)^2-\frac{3}{\left(\frac{\rho_{m-1}}{\tau_{m-1}}+\frac{1-\rho_{m+2}}{\tau_{m+1}}\right)^2}.
\end{align}
\end{widetext}

\section{Fr\'echet distribution}
\label{b}
Here, we derive that when random variables follow a power-law distribution (Eq.~(\ref{eq1})),
the maximum of those follows the Fr\'echet distribution using the extreme value theory \cite{HaanFerreira}.
We define $\tau_1, \dots,\tau_L$ as the random variables which follow the power-law distribution with exponent $\alpha$.
The probability for $\tau_m=\max\{\tau_1,\dots,\tau_L\}\leq s$ is given by
\begin{equation}
    \Pr(\tau_m\leq s)=\prod_{i=1}^L\Pr(\tau_i\leq s)=G(s)^L,
    \label{eqA1}
\end{equation}
where $G(s)=\Pr(\tau_i\leq s)=1-(s/\tau_c)^{-\alpha}$.
We normalize $\tau_m$ as
\begin{equation}
    X_\alpha=\frac{\tau_m}{\tau_cL^{1/\alpha}}
\end{equation}
for $L\rightarrow\infty$.
It follows that $\Pr(X_\alpha\leq x)=F_\alpha(x)$ is given by 
\begin{equation}
    F_\alpha(x)=\lim_{L\rightarrow\infty}G(\tau_cL^{1/\alpha}x)^L=\exp{(-x^{-\alpha})}.
\end{equation}
Therefore, the normalized $\tau_m$ follows the Fr\'echet distribution.
\bibliographystyle{apsrev4-1}
\bibliography{ASEP_D}

\begin{thebibliography}{48}%
\makeatletter
\providecommand \@ifxundefined [1]{%
 \@ifx{#1\undefined}
}%
\providecommand \@ifnum [1]{%
 \ifnum #1\expandafter \@firstoftwo
 \else \expandafter \@secondoftwo
 \fi
}%
\providecommand \@ifx [1]{%
 \ifx #1\expandafter \@firstoftwo
 \else \expandafter \@secondoftwo
 \fi
}%
\providecommand \natexlab [1]{#1}%
\providecommand \enquote  [1]{``#1''}%
\providecommand \bibnamefont  [1]{#1}%
\providecommand \bibfnamefont [1]{#1}%
\providecommand \citenamefont [1]{#1}%
\providecommand \href@noop [0]{\@secondoftwo}%
\providecommand \href [0]{\begingroup \@sanitize@url \@href}%
\providecommand \@href[1]{\@@startlink{#1}\@@href}%
\providecommand \@@href[1]{\endgroup#1\@@endlink}%
\providecommand \@sanitize@url [0]{\catcode `\\12\catcode `\$12\catcode
  `\&12\catcode `\#12\catcode `\^12\catcode `\_12\catcode `\%12\relax}%
\providecommand \@@startlink[1]{}%
\providecommand \@@endlink[0]{}%
\providecommand \url  [0]{\begingroup\@sanitize@url \@url }%
\providecommand \@url [1]{\endgroup\@href {#1}{\urlprefix }}%
\providecommand \urlprefix  [0]{URL }%
\providecommand \Eprint [0]{\href }%
\providecommand \doibase [0]{http://dx.doi.org/}%
\providecommand \selectlanguage [0]{\@gobble}%
\providecommand \bibinfo  [0]{\@secondoftwo}%
\providecommand \bibfield  [0]{\@secondoftwo}%
\providecommand \translation [1]{[#1]}%
\providecommand \BibitemOpen [0]{}%
\providecommand \bibitemStop [0]{}%
\providecommand \bibitemNoStop [0]{.\EOS\space}%
\providecommand \EOS [0]{\spacefactor3000\relax}%
\providecommand \BibitemShut  [1]{\csname bibitem#1\endcsname}%
\let\auto@bib@innerbib\@empty
\bibitem [{\citenamefont {Derrida}(1998)}]{Derrida}%
  \BibitemOpen
  \bibfield  {author} {\bibinfo {author} {\bibfnamefont {B.}~\bibnamefont
  {Derrida}},\ }\href {\doibase https://doi.org/10.1016/S0370-1573(98)00006-4}
  {\bibfield  {journal} {\bibinfo  {journal} {Phys. Rep.}\ }\textbf {\bibinfo
  {volume} {301}},\ \bibinfo {pages} {65} (\bibinfo {year} {1998})}\BibitemShut
  {NoStop}%
\bibitem [{\citenamefont {Arita}\ \emph {et~al.}(2017)\citenamefont {Arita},
  \citenamefont {Foulaadvand},\ and\ \citenamefont
  {Santen}}]{AritaFoulaadvandSanten}%
  \BibitemOpen
  \bibfield  {author} {\bibinfo {author} {\bibfnamefont {C.}~\bibnamefont
  {Arita}}, \bibinfo {author} {\bibfnamefont {M.~E.}\ \bibnamefont
  {Foulaadvand}}, \ and\ \bibinfo {author} {\bibfnamefont {L.}~\bibnamefont
  {Santen}},\ }\href {\doibase 10.1103/PhysRevE.95.032108} {\bibfield
  {journal} {\bibinfo  {journal} {Phys. Rev. E}\ }\textbf {\bibinfo {volume}
  {95}},\ \bibinfo {pages} {032108} (\bibinfo {year} {2017})}\BibitemShut
  {NoStop}%
\bibitem [{\citenamefont {Chou}\ and\ \citenamefont
  {Lakatos}(2004)}]{ChouLakatou}%
  \BibitemOpen
  \bibfield  {author} {\bibinfo {author} {\bibfnamefont {T.}~\bibnamefont
  {Chou}}\ and\ \bibinfo {author} {\bibfnamefont {G.}~\bibnamefont {Lakatos}},\
  }\href {\doibase 10.1103/PhysRevLett.93.198101} {\bibfield  {journal}
  {\bibinfo  {journal} {Phys. Rev. Lett.}\ }\textbf {\bibinfo {volume} {93}},\
  \bibinfo {pages} {198101} (\bibinfo {year} {2004})}\BibitemShut {NoStop}%
\bibitem [{\citenamefont {Ciandrini}\ \emph {et~al.}(2010)\citenamefont
  {Ciandrini}, \citenamefont {Stansfield},\ and\ \citenamefont
  {Romano}}]{CiandriniStansfieldRomano}%
  \BibitemOpen
  \bibfield  {author} {\bibinfo {author} {\bibfnamefont {L.}~\bibnamefont
  {Ciandrini}}, \bibinfo {author} {\bibfnamefont {I.}~\bibnamefont
  {Stansfield}}, \ and\ \bibinfo {author} {\bibfnamefont {M.~C.}\ \bibnamefont
  {Romano}},\ }\href {\doibase 10.1103/PhysRevE.81.051904} {\bibfield
  {journal} {\bibinfo  {journal} {Phys. Rev. E}\ }\textbf {\bibinfo {volume}
  {81}},\ \bibinfo {pages} {051904} (\bibinfo {year} {2010})}\BibitemShut
  {NoStop}%
\bibitem [{\citenamefont {Dana}\ and\ \citenamefont
  {Tuller}(2014)}]{DanaTuller}%
  \BibitemOpen
  \bibfield  {author} {\bibinfo {author} {\bibfnamefont {A.}~\bibnamefont
  {Dana}}\ and\ \bibinfo {author} {\bibfnamefont {T.}~\bibnamefont {Tuller}},\
  }\href {\doibase 10.1093/nar/gku646} {\bibfield  {journal} {\bibinfo
  {journal} {Nucleic Acids Res.}\ }\textbf {\bibinfo {volume} {42}},\ \bibinfo
  {pages} {9171} (\bibinfo {year} {2014})}\BibitemShut {NoStop}%
\bibitem [{\citenamefont {Kardar}\ \emph {et~al.}(1986)\citenamefont {Kardar},
  \citenamefont {Parisi},\ and\ \citenamefont {Zhang}}]{KardarParisiZhang}%
  \BibitemOpen
  \bibfield  {author} {\bibinfo {author} {\bibfnamefont {M.}~\bibnamefont
  {Kardar}}, \bibinfo {author} {\bibfnamefont {G.}~\bibnamefont {Parisi}}, \
  and\ \bibinfo {author} {\bibfnamefont {Y.-C.}\ \bibnamefont {Zhang}},\ }\href
  {\doibase 10.1103/PhysRevLett.56.889} {\bibfield  {journal} {\bibinfo
  {journal} {Phys. Rev. Lett.}\ }\textbf {\bibinfo {volume} {56}},\ \bibinfo
  {pages} {889} (\bibinfo {year} {1986})}\BibitemShut {NoStop}%
\bibitem [{\citenamefont {Johansson}(2000)}]{Johansson:2000tk}%
  \BibitemOpen
  \bibfield  {author} {\bibinfo {author} {\bibfnamefont {K.}~\bibnamefont
  {Johansson}},\ }\href {\doibase 10.1007/s002200050027} {\bibfield  {journal}
  {\bibinfo  {journal} {Comm. Math. Phys.}\ }\textbf {\bibinfo {volume}
  {209}},\ \bibinfo {pages} {437} (\bibinfo {year} {2000})}\BibitemShut
  {NoStop}%
\bibitem [{\citenamefont {Tracy}\ and\ \citenamefont
  {Widom}(2009)}]{Tracy:2009tx}%
  \BibitemOpen
  \bibfield  {author} {\bibinfo {author} {\bibfnamefont {C.~A.}\ \bibnamefont
  {Tracy}}\ and\ \bibinfo {author} {\bibfnamefont {H.}~\bibnamefont {Widom}},\
  }\href {\doibase 10.1007/s00220-009-0761-0} {\bibfield  {journal} {\bibinfo
  {journal} {Comm. Math. Phys.}\ }\textbf {\bibinfo {volume} {290}},\ \bibinfo
  {pages} {129} (\bibinfo {year} {2009})}\BibitemShut {NoStop}%
\bibitem [{\citenamefont {Aggarwal}(2018)}]{Aggarwal}%
  \BibitemOpen
  \bibfield  {author} {\bibinfo {author} {\bibfnamefont {A.}~\bibnamefont
  {Aggarwal}},\ }\href {\doibase 10.1215/00127094-2017-0029} {\bibfield
  {journal} {\bibinfo  {journal} {Duke Math. J.}\ }\textbf {\bibinfo {volume}
  {167}},\ \bibinfo {pages} {269 } (\bibinfo {year} {2018})}\BibitemShut
  {NoStop}%
\bibitem [{\citenamefont {Sasamoto}\ and\ \citenamefont
  {Spohn}(2010{\natexlab{a}})}]{PhysRevLett.104.230602}%
  \BibitemOpen
  \bibfield  {author} {\bibinfo {author} {\bibfnamefont {T.}~\bibnamefont
  {Sasamoto}}\ and\ \bibinfo {author} {\bibfnamefont {H.}~\bibnamefont
  {Spohn}},\ }\href {\doibase 10.1103/PhysRevLett.104.230602} {\bibfield
  {journal} {\bibinfo  {journal} {Phys. Rev. Lett.}\ }\textbf {\bibinfo
  {volume} {104}},\ \bibinfo {pages} {230602} (\bibinfo {year}
  {2010}{\natexlab{a}})}\BibitemShut {NoStop}%
\bibitem [{\citenamefont {Sasamoto}\ and\ \citenamefont
  {Spohn}(2010{\natexlab{b}})}]{SASAMOTO2010523}%
  \BibitemOpen
  \bibfield  {author} {\bibinfo {author} {\bibfnamefont {T.}~\bibnamefont
  {Sasamoto}}\ and\ \bibinfo {author} {\bibfnamefont {H.}~\bibnamefont
  {Spohn}},\ }\href {\doibase https://doi.org/10.1016/j.nuclphysb.2010.03.026}
  {\bibfield  {journal} {\bibinfo  {journal} {Nuclear Phys. B}\ }\textbf
  {\bibinfo {volume} {834}},\ \bibinfo {pages} {523} (\bibinfo {year}
  {2010}{\natexlab{b}})}\BibitemShut {NoStop}%
\bibitem [{\citenamefont {Amir}\ \emph {et~al.}(2011)\citenamefont {Amir},
  \citenamefont {Corwin},\ and\ \citenamefont {Quastel}}]{AmirGideon}%
  \BibitemOpen
  \bibfield  {author} {\bibinfo {author} {\bibfnamefont {G.}~\bibnamefont
  {Amir}}, \bibinfo {author} {\bibfnamefont {I.}~\bibnamefont {Corwin}}, \ and\
  \bibinfo {author} {\bibfnamefont {J.}~\bibnamefont {Quastel}},\ }\href
  {\doibase https://doi.org/10.1002/cpa.20347} {\bibfield  {journal} {\bibinfo
  {journal} {Comm. Pure Appl. Math.}\ }\textbf {\bibinfo {volume} {64}},\
  \bibinfo {pages} {466} (\bibinfo {year} {2011})}\BibitemShut {NoStop}%
\bibitem [{\citenamefont {Derrida}\ and\ \citenamefont
  {Lebowitz}(1998)}]{PhysRevLett.80.209}%
  \BibitemOpen
  \bibfield  {author} {\bibinfo {author} {\bibfnamefont {B.}~\bibnamefont
  {Derrida}}\ and\ \bibinfo {author} {\bibfnamefont {J.~L.}\ \bibnamefont
  {Lebowitz}},\ }\href {\doibase 10.1103/PhysRevLett.80.209} {\bibfield
  {journal} {\bibinfo  {journal} {Phys. Rev. Lett.}\ }\textbf {\bibinfo
  {volume} {80}},\ \bibinfo {pages} {209} (\bibinfo {year} {1998})}\BibitemShut
  {NoStop}%
\bibitem [{\citenamefont {Bertini}\ \emph {et~al.}(2005)\citenamefont
  {Bertini}, \citenamefont {De~Sole}, \citenamefont {Gabrielli}, \citenamefont
  {Jona-Lasinio},\ and\ \citenamefont {Landim}}]{PhysRevLett.94.030601}%
  \BibitemOpen
  \bibfield  {author} {\bibinfo {author} {\bibfnamefont {L.}~\bibnamefont
  {Bertini}}, \bibinfo {author} {\bibfnamefont {A.}~\bibnamefont {De~Sole}},
  \bibinfo {author} {\bibfnamefont {D.}~\bibnamefont {Gabrielli}}, \bibinfo
  {author} {\bibfnamefont {G.}~\bibnamefont {Jona-Lasinio}}, \ and\ \bibinfo
  {author} {\bibfnamefont {C.}~\bibnamefont {Landim}},\ }\href {\doibase
  10.1103/PhysRevLett.94.030601} {\bibfield  {journal} {\bibinfo  {journal}
  {Phys. Rev. Lett.}\ }\textbf {\bibinfo {volume} {94}},\ \bibinfo {pages}
  {030601} (\bibinfo {year} {2005})}\BibitemShut {NoStop}%
\bibitem [{\citenamefont {Lips}\ \emph {et~al.}(2018)\citenamefont {Lips},
  \citenamefont {Ryabov},\ and\ \citenamefont {Maass}}]{LipsRyabovMaass}%
  \BibitemOpen
  \bibfield  {author} {\bibinfo {author} {\bibfnamefont {D.}~\bibnamefont
  {Lips}}, \bibinfo {author} {\bibfnamefont {A.}~\bibnamefont {Ryabov}}, \ and\
  \bibinfo {author} {\bibfnamefont {P.}~\bibnamefont {Maass}},\ }\href
  {\doibase 10.1103/PhysRevLett.121.160601} {\bibfield  {journal} {\bibinfo
  {journal} {Phys. Rev. Lett.}\ }\textbf {\bibinfo {volume} {121}},\ \bibinfo
  {pages} {160601} (\bibinfo {year} {2018})}\BibitemShut {NoStop}%
\bibitem [{\citenamefont {Concannon}\ and\ \citenamefont
  {Blythe}(2014)}]{ConcannonBlythe}%
  \BibitemOpen
  \bibfield  {author} {\bibinfo {author} {\bibfnamefont {R.~J.}\ \bibnamefont
  {Concannon}}\ and\ \bibinfo {author} {\bibfnamefont {R.~A.}\ \bibnamefont
  {Blythe}},\ }\href {\doibase 10.1103/PhysRevLett.112.050603} {\bibfield
  {journal} {\bibinfo  {journal} {Phys. Rev. Lett.}\ }\textbf {\bibinfo
  {volume} {112}},\ \bibinfo {pages} {050603} (\bibinfo {year}
  {2014})}\BibitemShut {NoStop}%
\bibitem [{\citenamefont {Tripathy}\ and\ \citenamefont
  {Barma}(1998)}]{TripathyBarma}%
  \BibitemOpen
  \bibfield  {author} {\bibinfo {author} {\bibfnamefont {G.}~\bibnamefont
  {Tripathy}}\ and\ \bibinfo {author} {\bibfnamefont {M.}~\bibnamefont
  {Barma}},\ }\href {\doibase 10.1103/PhysRevE.58.1911} {\bibfield  {journal}
  {\bibinfo  {journal} {Phys. Rev. E}\ }\textbf {\bibinfo {volume} {58}},\
  \bibinfo {pages} {1911} (\bibinfo {year} {1998})}\BibitemShut {NoStop}%
\bibitem [{\citenamefont {Enaud}\ and\ \citenamefont
  {Derrida}(2004)}]{Enaud_2004}%
  \BibitemOpen
  \bibfield  {author} {\bibinfo {author} {\bibfnamefont {C.}~\bibnamefont
  {Enaud}}\ and\ \bibinfo {author} {\bibfnamefont {B.}~\bibnamefont
  {Derrida}},\ }\href {\doibase 10.1209/epl/i2003-10153-8} {\bibfield
  {journal} {\bibinfo  {journal} {Europhys. Lett.}\ }\textbf {\bibinfo {volume}
  {66}},\ \bibinfo {pages} {83} (\bibinfo {year} {2004})}\BibitemShut {NoStop}%
\bibitem [{\citenamefont {Harris}\ and\ \citenamefont
  {Stinchcombe}(2004)}]{HarrisStinchcombe}%
  \BibitemOpen
  \bibfield  {author} {\bibinfo {author} {\bibfnamefont {R.~J.}\ \bibnamefont
  {Harris}}\ and\ \bibinfo {author} {\bibfnamefont {R.~B.}\ \bibnamefont
  {Stinchcombe}},\ }\href {\doibase 10.1103/PhysRevE.70.016108} {\bibfield
  {journal} {\bibinfo  {journal} {Phys. Rev. E}\ }\textbf {\bibinfo {volume}
  {70}},\ \bibinfo {pages} {016108} (\bibinfo {year} {2004})}\BibitemShut
  {NoStop}%
\bibitem [{\citenamefont {Juh\'asz}\ \emph {et~al.}(2006)\citenamefont
  {Juh\'asz}, \citenamefont {Santen},\ and\ \citenamefont
  {Igl\'oi}}]{JuhaszSantenIgloi}%
  \BibitemOpen
  \bibfield  {author} {\bibinfo {author} {\bibfnamefont {R.}~\bibnamefont
  {Juh\'asz}}, \bibinfo {author} {\bibfnamefont {L.}~\bibnamefont {Santen}}, \
  and\ \bibinfo {author} {\bibfnamefont {F.}~\bibnamefont {Igl\'oi}},\ }\href
  {\doibase 10.1103/PhysRevE.74.061101} {\bibfield  {journal} {\bibinfo
  {journal} {Phys. Rev. E}\ }\textbf {\bibinfo {volume} {74}},\ \bibinfo
  {pages} {061101} (\bibinfo {year} {2006})}\BibitemShut {NoStop}%
\bibitem [{\citenamefont {Stinchcombe}\ and\ \citenamefont
  {de~Queiroz}(2011)}]{StinchcombeQueiroz}%
  \BibitemOpen
  \bibfield  {author} {\bibinfo {author} {\bibfnamefont {R.~B.}\ \bibnamefont
  {Stinchcombe}}\ and\ \bibinfo {author} {\bibfnamefont {S.~L.~A.}\
  \bibnamefont {de~Queiroz}},\ }\href {\doibase 10.1103/PhysRevE.83.061113}
  {\bibfield  {journal} {\bibinfo  {journal} {Phys. Rev. E}\ }\textbf {\bibinfo
  {volume} {83}},\ \bibinfo {pages} {061113} (\bibinfo {year}
  {2011})}\BibitemShut {NoStop}%
\bibitem [{\citenamefont {Nossan}(2013)}]{Nossan_2013}%
  \BibitemOpen
  \bibfield  {author} {\bibinfo {author} {\bibfnamefont {J.~S.}\ \bibnamefont
  {Nossan}},\ }\href {\doibase 10.1088/1751-8113/46/31/315001} {\bibfield
  {journal} {\bibinfo  {journal} {J. Phys. A: Math. Theor.}\ }\textbf {\bibinfo
  {volume} {46}},\ \bibinfo {pages} {315001} (\bibinfo {year}
  {2013})}\BibitemShut {NoStop}%
\bibitem [{\citenamefont {Bahadoran}\ and\ \citenamefont
  {Bodineau}(2015)}]{10.1214/14-BJPS277}%
  \BibitemOpen
  \bibfield  {author} {\bibinfo {author} {\bibfnamefont {C.}~\bibnamefont
  {Bahadoran}}\ and\ \bibinfo {author} {\bibfnamefont {T.}~\bibnamefont
  {Bodineau}},\ }\href {\doibase 10.1214/14-BJPS277} {\bibfield  {journal}
  {\bibinfo  {journal} {Braz. J. Probab. Stat.}\ }\textbf {\bibinfo {volume}
  {29}},\ \bibinfo {pages} {282 } (\bibinfo {year} {2015})}\BibitemShut
  {NoStop}%
\bibitem [{\citenamefont {Banerjee}\ and\ \citenamefont
  {Basu}(2020)}]{BanerjeeBasu}%
  \BibitemOpen
  \bibfield  {author} {\bibinfo {author} {\bibfnamefont {T.}~\bibnamefont
  {Banerjee}}\ and\ \bibinfo {author} {\bibfnamefont {A.}~\bibnamefont
  {Basu}},\ }\href {\doibase 10.1103/PhysRevResearch.2.013025} {\bibfield
  {journal} {\bibinfo  {journal} {Phys. Rev. Res.}\ }\textbf {\bibinfo {volume}
  {2}},\ \bibinfo {pages} {013025} (\bibinfo {year} {2020})}\BibitemShut
  {NoStop}%
\bibitem [{\citenamefont {Haldar}\ and\ \citenamefont
  {Basu}(2020)}]{PhysRevResearch.2.043073}%
  \BibitemOpen
  \bibfield  {author} {\bibinfo {author} {\bibfnamefont {A.}~\bibnamefont
  {Haldar}}\ and\ \bibinfo {author} {\bibfnamefont {A.}~\bibnamefont {Basu}},\
  }\href {\doibase 10.1103/PhysRevResearch.2.043073} {\bibfield  {journal}
  {\bibinfo  {journal} {Phys. Rev. Res.}\ }\textbf {\bibinfo {volume} {2}},\
  \bibinfo {pages} {043073} (\bibinfo {year} {2020})}\BibitemShut {NoStop}%
\bibitem [{\citenamefont {Derrida}\ \emph {et~al.}(1993)\citenamefont
  {Derrida}, \citenamefont {Evans},\ and\ \citenamefont
  {Mukamel}}]{Derrida_1993}%
  \BibitemOpen
  \bibfield  {author} {\bibinfo {author} {\bibfnamefont {B.}~\bibnamefont
  {Derrida}}, \bibinfo {author} {\bibfnamefont {M.~R.}\ \bibnamefont {Evans}},
  \ and\ \bibinfo {author} {\bibfnamefont {D.}~\bibnamefont {Mukamel}},\ }\href
  {\doibase 10.1088/0305-4470/26/19/023} {\bibfield  {journal} {\bibinfo
  {journal} {J. Phys. A: Math. Gen.}\ }\textbf {\bibinfo {volume} {26}},\
  \bibinfo {pages} {4911} (\bibinfo {year} {1993})}\BibitemShut {NoStop}%
\bibitem [{\citenamefont {Neri}\ \emph {et~al.}(2011)\citenamefont {Neri},
  \citenamefont {Kern},\ and\ \citenamefont
  {Parmeggiani}}]{PhysRevLett.107.068702}%
  \BibitemOpen
  \bibfield  {author} {\bibinfo {author} {\bibfnamefont {I.}~\bibnamefont
  {Neri}}, \bibinfo {author} {\bibfnamefont {N.}~\bibnamefont {Kern}}, \ and\
  \bibinfo {author} {\bibfnamefont {A.}~\bibnamefont {Parmeggiani}},\ }\href
  {\doibase 10.1103/PhysRevLett.107.068702} {\bibfield  {journal} {\bibinfo
  {journal} {Phys. Rev. Lett.}\ }\textbf {\bibinfo {volume} {107}},\ \bibinfo
  {pages} {068702} (\bibinfo {year} {2011})}\BibitemShut {NoStop}%
\bibitem [{\citenamefont {Neri}\ \emph {et~al.}(2013)\citenamefont {Neri},
  \citenamefont {Kern},\ and\ \citenamefont {Parmeggiani}}]{Neri_2013}%
  \BibitemOpen
  \bibfield  {author} {\bibinfo {author} {\bibfnamefont {I.}~\bibnamefont
  {Neri}}, \bibinfo {author} {\bibfnamefont {N.}~\bibnamefont {Kern}}, \ and\
  \bibinfo {author} {\bibfnamefont {A.}~\bibnamefont {Parmeggiani}},\ }\href
  {\doibase 10.1088/1367-2630/15/8/085005} {\bibfield  {journal} {\bibinfo
  {journal} {New J. Phys.}\ }\textbf {\bibinfo {volume} {15}},\ \bibinfo
  {pages} {085005} (\bibinfo {year} {2013})}\BibitemShut {NoStop}%
\bibitem [{\citenamefont {Denisov}\ \emph {et~al.}(2015)\citenamefont
  {Denisov}, \citenamefont {Miedema}, \citenamefont {Nienhuis},\ and\
  \citenamefont {Schall}}]{PhysRevE.92.052714}%
  \BibitemOpen
  \bibfield  {author} {\bibinfo {author} {\bibfnamefont {D.~V.}\ \bibnamefont
  {Denisov}}, \bibinfo {author} {\bibfnamefont {D.~M.}\ \bibnamefont
  {Miedema}}, \bibinfo {author} {\bibfnamefont {B.}~\bibnamefont {Nienhuis}}, \
  and\ \bibinfo {author} {\bibfnamefont {P.}~\bibnamefont {Schall}},\ }\href
  {\doibase 10.1103/PhysRevE.92.052714} {\bibfield  {journal} {\bibinfo
  {journal} {Phys. Rev. E}\ }\textbf {\bibinfo {volume} {92}},\ \bibinfo
  {pages} {052714} (\bibinfo {year} {2015})}\BibitemShut {NoStop}%
\bibitem [{\citenamefont {Metzler}\ and\ \citenamefont
  {Klafter}(2000)}]{METZLER20001}%
  \BibitemOpen
  \bibfield  {author} {\bibinfo {author} {\bibfnamefont {R.}~\bibnamefont
  {Metzler}}\ and\ \bibinfo {author} {\bibfnamefont {J.}~\bibnamefont
  {Klafter}},\ }\href {\doibase https://doi.org/10.1016/S0370-1573(00)00070-3}
  {\bibfield  {journal} {\bibinfo  {journal} {Phys. Rep.}\ }\textbf {\bibinfo
  {volume} {339}},\ \bibinfo {pages} {1} (\bibinfo {year} {2000})}\BibitemShut
  {NoStop}%
\bibitem [{\citenamefont {Bouchaud}\ and\ \citenamefont
  {Georges}(1990)}]{BouchaudGeorfes}%
  \BibitemOpen
  \bibfield  {author} {\bibinfo {author} {\bibfnamefont {J.~P.}\ \bibnamefont
  {Bouchaud}}\ and\ \bibinfo {author} {\bibfnamefont {A.}~\bibnamefont
  {Georges}},\ }\href@noop {} {\bibfield  {journal} {\bibinfo  {journal} {Phys.
  Rep.}\ }\textbf {\bibinfo {volume} {195}} (\bibinfo {year}
  {1990})}\BibitemShut {NoStop}%
\bibitem [{\citenamefont {Akimoto}\ \emph {et~al.}(2016)\citenamefont
  {Akimoto}, \citenamefont {Barkai},\ and\ \citenamefont
  {Saito}}]{AkimotoBarkaiSaito}%
  \BibitemOpen
  \bibfield  {author} {\bibinfo {author} {\bibfnamefont {T.}~\bibnamefont
  {Akimoto}}, \bibinfo {author} {\bibfnamefont {E.}~\bibnamefont {Barkai}}, \
  and\ \bibinfo {author} {\bibfnamefont {K.}~\bibnamefont {Saito}},\ }\href
  {\doibase 10.1103/PhysRevLett.117.180602} {\bibfield  {journal} {\bibinfo
  {journal} {Phys. Rev. Lett.}\ }\textbf {\bibinfo {volume} {117}},\ \bibinfo
  {pages} {180602} (\bibinfo {year} {2016})}\BibitemShut {NoStop}%
\bibitem [{\citenamefont {Akimoto}\ \emph {et~al.}(2018)\citenamefont
  {Akimoto}, \citenamefont {Barkai},\ and\ \citenamefont
  {Saito}}]{AkimotoBarkaiSaito2018}%
  \BibitemOpen
  \bibfield  {author} {\bibinfo {author} {\bibfnamefont {T.}~\bibnamefont
  {Akimoto}}, \bibinfo {author} {\bibfnamefont {E.}~\bibnamefont {Barkai}}, \
  and\ \bibinfo {author} {\bibfnamefont {K.}~\bibnamefont {Saito}},\ }\href
  {\doibase 10.1103/PhysRevE.97.052143} {\bibfield  {journal} {\bibinfo
  {journal} {Phys. Rev. E}\ }\textbf {\bibinfo {volume} {97}},\ \bibinfo
  {pages} {052143} (\bibinfo {year} {2018})}\BibitemShut {NoStop}%
\bibitem [{\citenamefont {Luo}\ and\ \citenamefont {Yi}(2018)}]{LuoYi}%
  \BibitemOpen
  \bibfield  {author} {\bibinfo {author} {\bibfnamefont {L.}~\bibnamefont
  {Luo}}\ and\ \bibinfo {author} {\bibfnamefont {M.}~\bibnamefont {Yi}},\
  }\href {\doibase 10.1103/PhysRevE.97.042122} {\bibfield  {journal} {\bibinfo
  {journal} {Phys. Rev. E}\ }\textbf {\bibinfo {volume} {97}},\ \bibinfo
  {pages} {042122} (\bibinfo {year} {2018})}\BibitemShut {NoStop}%
\bibitem [{\citenamefont {Akimoto}\ and\ \citenamefont
  {Saito}(2020)}]{AkimotoSaito2020}%
  \BibitemOpen
  \bibfield  {author} {\bibinfo {author} {\bibfnamefont {T.}~\bibnamefont
  {Akimoto}}\ and\ \bibinfo {author} {\bibfnamefont {K.}~\bibnamefont
  {Saito}},\ }\href {\doibase 10.1103/PhysRevE.101.042133} {\bibfield
  {journal} {\bibinfo  {journal} {Phys. Rev. E}\ }\textbf {\bibinfo {volume}
  {101}},\ \bibinfo {pages} {042133} (\bibinfo {year} {2020})}\BibitemShut
  {NoStop}%
\bibitem [{\citenamefont {Metzler}\ \emph {et~al.}(2014)\citenamefont
  {Metzler}, \citenamefont {Sanders}, \citenamefont {Lomholt}, \citenamefont
  {Lizana}, \citenamefont {Fogelmark},\ and\ \citenamefont
  {Ambj{\"o}rnsson}}]{Metzler:2014aa}%
  \BibitemOpen
  \bibfield  {author} {\bibinfo {author} {\bibfnamefont {R.}~\bibnamefont
  {Metzler}}, \bibinfo {author} {\bibfnamefont {L.}~\bibnamefont {Sanders}},
  \bibinfo {author} {\bibfnamefont {M.~A.}\ \bibnamefont {Lomholt}}, \bibinfo
  {author} {\bibfnamefont {L.}~\bibnamefont {Lizana}}, \bibinfo {author}
  {\bibfnamefont {K.}~\bibnamefont {Fogelmark}}, \ and\ \bibinfo {author}
  {\bibfnamefont {T.}~\bibnamefont {Ambj{\"o}rnsson}},\ }\href {\doibase
  10.1140/epjst/e2014-02333-5} {\bibfield  {journal} {\bibinfo  {journal} {EPJ
  Special Topics}\ }\textbf {\bibinfo {volume} {223}},\ \bibinfo {pages} {3287}
  (\bibinfo {year} {2014})}\BibitemShut {NoStop}%
\bibitem [{\citenamefont {Sanders}\ \emph {et~al.}(2014)\citenamefont
  {Sanders}, \citenamefont {Lomholt}, \citenamefont {Lizana}, \citenamefont
  {Fogelmark}, \citenamefont {Metzler},\ and\ \citenamefont
  {Ambj{\"o}rnsson}}]{Sanders_2014}%
  \BibitemOpen
  \bibfield  {author} {\bibinfo {author} {\bibfnamefont {L.~P.}\ \bibnamefont
  {Sanders}}, \bibinfo {author} {\bibfnamefont {M.~A.}\ \bibnamefont
  {Lomholt}}, \bibinfo {author} {\bibfnamefont {L.}~\bibnamefont {Lizana}},
  \bibinfo {author} {\bibfnamefont {K.}~\bibnamefont {Fogelmark}}, \bibinfo
  {author} {\bibfnamefont {R.}~\bibnamefont {Metzler}}, \ and\ \bibinfo
  {author} {\bibfnamefont {T.}~\bibnamefont {Ambj{\"o}rnsson}},\ }\href
  {\doibase 10.1088/1367-2630/16/11/113050} {\bibfield  {journal} {\bibinfo
  {journal} {New J. Phys.}\ }\textbf {\bibinfo {volume} {16}},\ \bibinfo
  {pages} {113050} (\bibinfo {year} {2014})}\BibitemShut {NoStop}%
\bibitem [{\citenamefont {Gran{\'e}li}\ \emph {et~al.}(2006)\citenamefont
  {Gran{\'e}li}, \citenamefont {Yeykal}, \citenamefont {Robertson},\ and\
  \citenamefont {Greene}}]{GraneliGreeneRobertsonYeykal}%
  \BibitemOpen
  \bibfield  {author} {\bibinfo {author} {\bibfnamefont {A.}~\bibnamefont
  {Gran{\'e}li}}, \bibinfo {author} {\bibfnamefont {C.~C.}\ \bibnamefont
  {Yeykal}}, \bibinfo {author} {\bibfnamefont {R.~B.}\ \bibnamefont
  {Robertson}}, \ and\ \bibinfo {author} {\bibfnamefont {E.~C.}\ \bibnamefont
  {Greene}},\ }\href {\doibase 10.1073/pnas.0508366103} {\bibfield  {journal}
  {\bibinfo  {journal} {Proc. Natl. Acad. Sci. U.S.A.}\ }\textbf {\bibinfo
  {volume} {103}},\ \bibinfo {pages} {1221} (\bibinfo {year}
  {2006})}\BibitemShut {NoStop}%
\bibitem [{\citenamefont {Wang}\ \emph {et~al.}(2006)\citenamefont {Wang},
  \citenamefont {Austin},\ and\ \citenamefont {Cox}}]{AustinCoxWang}%
  \BibitemOpen
  \bibfield  {author} {\bibinfo {author} {\bibfnamefont {Y.~M.}\ \bibnamefont
  {Wang}}, \bibinfo {author} {\bibfnamefont {R.~H.}\ \bibnamefont {Austin}}, \
  and\ \bibinfo {author} {\bibfnamefont {E.~C.}\ \bibnamefont {Cox}},\ }\href
  {\doibase 10.1103/PhysRevLett.97.048302} {\bibfield  {journal} {\bibinfo
  {journal} {Phys. Rev. Lett.}\ }\textbf {\bibinfo {volume} {97}},\ \bibinfo
  {pages} {048302} (\bibinfo {year} {2006})}\BibitemShut {NoStop}%
\bibitem [{\citenamefont {Yamamoto}\ \emph {et~al.}(2014)\citenamefont
  {Yamamoto}, \citenamefont {Akimoto}, \citenamefont {Hirano}, \citenamefont
  {Yasui},\ and\ \citenamefont {Yasuoka}}]{AkimotoHiraoYamamotoYasuiYasuoka}%
  \BibitemOpen
  \bibfield  {author} {\bibinfo {author} {\bibfnamefont {E.}~\bibnamefont
  {Yamamoto}}, \bibinfo {author} {\bibfnamefont {T.}~\bibnamefont {Akimoto}},
  \bibinfo {author} {\bibfnamefont {Y.}~\bibnamefont {Hirano}}, \bibinfo
  {author} {\bibfnamefont {M.}~\bibnamefont {Yasui}}, \ and\ \bibinfo {author}
  {\bibfnamefont {K.}~\bibnamefont {Yasuoka}},\ }\href {\doibase
  10.1103/PhysRevE.89.022718} {\bibfield  {journal} {\bibinfo  {journal} {Phys.
  Rev. E}\ }\textbf {\bibinfo {volume} {89}},\ \bibinfo {pages} {022718}
  (\bibinfo {year} {2014})}\BibitemShut {NoStop}%
\bibitem [{\citenamefont {Sakai}\ and\ \citenamefont
  {Akimoto}(2022)}]{https://doi.org/10.48550/arxiv.2208.10102}%
  \BibitemOpen
  \bibfield  {author} {\bibinfo {author} {\bibfnamefont {I.}~\bibnamefont
  {Sakai}}\ and\ \bibinfo {author} {\bibfnamefont {T.}~\bibnamefont
  {Akimoto}},\ }\href {https://arxiv.org/abs/2208.10102} {\bibfield  {journal}
  {\bibinfo  {journal} {arXiv:2208.10102}\ } (\bibinfo {year}
  {2022})}\BibitemShut {NoStop}%
\bibitem [{\citenamefont {Akimoto}\ and\ \citenamefont
  {Saito}(2019)}]{AkimotoSaito2019}%
  \BibitemOpen
  \bibfield  {author} {\bibinfo {author} {\bibfnamefont {T.}~\bibnamefont
  {Akimoto}}\ and\ \bibinfo {author} {\bibfnamefont {K.}~\bibnamefont
  {Saito}},\ }\href {\doibase 10.1103/PhysRevE.99.052127} {\bibfield  {journal}
  {\bibinfo  {journal} {Phys. Rev. E}\ }\textbf {\bibinfo {volume} {99}},\
  \bibinfo {pages} {052127} (\bibinfo {year} {2019})}\BibitemShut {NoStop}%
\bibitem [{\citenamefont {Gupta}\ \emph {et~al.}(2007)\citenamefont {Gupta},
  \citenamefont {Majumdar}, \citenamefont {Godr\`eche},\ and\ \citenamefont
  {Barma}}]{GuptaMajumdarGodrecheBarma}%
  \BibitemOpen
  \bibfield  {author} {\bibinfo {author} {\bibfnamefont {S.}~\bibnamefont
  {Gupta}}, \bibinfo {author} {\bibfnamefont {S.~N.}\ \bibnamefont {Majumdar}},
  \bibinfo {author} {\bibfnamefont {C.}~\bibnamefont {Godr\`eche}}, \ and\
  \bibinfo {author} {\bibfnamefont {M.}~\bibnamefont {Barma}},\ }\href
  {\doibase 10.1103/PhysRevE.76.021112} {\bibfield  {journal} {\bibinfo
  {journal} {Phys. Rev. E}\ }\textbf {\bibinfo {volume} {76}},\ \bibinfo
  {pages} {021112} (\bibinfo {year} {2007})}\BibitemShut {NoStop}%
\bibitem [{\citenamefont {Janowsky}\ and\ \citenamefont
  {Lebowitz}(1992)}]{JanowskyLebowitz}%
  \BibitemOpen
  \bibfield  {author} {\bibinfo {author} {\bibfnamefont {S.~A.}\ \bibnamefont
  {Janowsky}}\ and\ \bibinfo {author} {\bibfnamefont {J.~L.}\ \bibnamefont
  {Lebowitz}},\ }\href {\doibase 10.1103/PhysRevA.45.618} {\bibfield  {journal}
  {\bibinfo  {journal} {Phys. Rev. A}\ }\textbf {\bibinfo {volume} {45}},\
  \bibinfo {pages} {618} (\bibinfo {year} {1992})}\BibitemShut {NoStop}%
\bibitem [{\citenamefont {de~Haan}\ and\ \citenamefont
  {Ferreira}(2006)}]{HaanFerreira}%
  \BibitemOpen
  \bibfield  {author} {\bibinfo {author} {\bibfnamefont {L.}~\bibnamefont
  {de~Haan}}\ and\ \bibinfo {author} {\bibfnamefont {A.}~\bibnamefont
  {Ferreira}},\ }\href@noop {} {\emph {\bibinfo {title} {Extreme value theory:
  an introduction}}},\ Vol.~\bibinfo {volume} {21}\ (\bibinfo  {publisher}
  {Springer},\ \bibinfo {year} {2006})\BibitemShut {NoStop}%
\bibitem [{\citenamefont {Godr{\`e}che}\ and\ \citenamefont
  {Luck}(2001)}]{Godreche}%
  \BibitemOpen
  \bibfield  {author} {\bibinfo {author} {\bibfnamefont {C.}~\bibnamefont
  {Godr{\`e}che}}\ and\ \bibinfo {author} {\bibfnamefont {J.~M.}\ \bibnamefont
  {Luck}},\ }\href {\doibase 10.1023/A:1010364003250} {\bibfield  {journal}
  {\bibinfo  {journal} {J. Stat. Phys.}\ }\textbf {\bibinfo {volume} {104}},\
  \bibinfo {pages} {489} (\bibinfo {year} {2001})}\BibitemShut {NoStop}%
\bibitem [{\citenamefont {Feller}(1971)}]{Feller1971}%
  \BibitemOpen
  \bibfield  {author} {\bibinfo {author} {\bibfnamefont {W.}~\bibnamefont
  {Feller}},\ }\href@noop {} {\emph {\bibinfo {title} {An Introduction to
  Probability Theory and its Applications}}},\ \bibinfo {edition} {2nd}\ ed.,\
  Vol.~\bibinfo {volume} {2}\ (\bibinfo  {publisher} {Wiley, New York},\
  \bibinfo {year} {1971})\BibitemShut {NoStop}%
\bibitem [{\citenamefont {van Beijeren}(1991)}]{Beijeren:1991aa}%
  \BibitemOpen
  \bibfield  {author} {\bibinfo {author} {\bibfnamefont {H.}~\bibnamefont {van
  Beijeren}},\ }\href {\doibase 10.1007/BF01026591} {\bibfield  {journal}
  {\bibinfo  {journal} {J. Stat. Phys.}\ }\textbf {\bibinfo {volume} {63}},\
  \bibinfo {pages} {47} (\bibinfo {year} {1991})}\BibitemShut {NoStop}%
\end{thebibliography}%

\end{document}